\documentclass[10pt]{article}
\pdfoutput=1
\usepackage{amsmath}
\usepackage{amssymb}
\usepackage{mathtools}
\usepackage{mathrsfs}
\usepackage{graphicx}
\usepackage{caption}
\usepackage{subcaption}
\usepackage{pdflscape}
\usepackage{pdfpages}
\usepackage{enumerate}
\usepackage{array}
\usepackage{makecell}
\usepackage{ctable}
\usepackage{morefloats}
\usepackage{setspace}
\usepackage{authblk}
\usepackage[]{appendix}
\usepackage{cite}
\usepackage{tabu}
\usepackage{geometry}
\usepackage[parfill]{parskip}
\usepackage[hidelinks]{hyperref}
\usepackage{titlesec}

\titleformat*{\section}{\large\bfseries}
\titleformat*{\subsection}{\normalsize\bfseries}
\titleformat*{\subsubsection}{\small\bfseries}

\begin{document}

\title{A Heuristic Reference Recursive Recipe for the Menacing Problem of Adaptively Tuning the Kalman Filter Statistics.\\ Part-2. Real Data Studies.}
\author{Shyam Mohan M$^1$ \footnote{shyammoh.2014@iitkalumni.org} , Naren Naik$^2$ \footnote{nnaik@iitk.ac.in} , R. M. O. Gemson$^3$ \footnote{mogratnam@rediffmail.com} , M. R. Ananthasayanam$^4$ \footnote{sayanam2005@yahoo.co.in (corresponding author)}\\
{\normalsize $^1$Formerly Post Graduate Student, IIT, Kanpur, India}\\
{\normalsize $^2$Department of Electrical Engineering, IIT, Kanpur, India}\\
{\normalsize $^3$Formerly Additional General Manager, HAL, Bangalore, India}\\
{\normalsize $^4$Formerly Professor, Department of Aerospace Engineering, IISc, Banglore}}
\date{}
\maketitle

\textbf{Abstract.} In the previous paper an adaptive filtering based on a reference recursive recipe was developed and tested on a simulated dynamics of a spring, mass, and damper with a weak nonlinear spring. In this paper the above recipe is applied to a more involved case of three sets of airplane data which have a larger number of state, measurements, and unknown parameters. Further the flight tests cannot always be conducted in an ideal situation of the process noise and the measurement noises being white and Gaussian as is generally assumed in the Kalman filter. The measurements are not available in general with respect to the center of gravity, possess scale and bias factors which will have to be modelled and estimated as well. The coupling between the longitudinal and lateral motion brings in added difficulty but makes the problem more interesting. At times the noisy measurements from the longitudinal and lateral motion are input into the longitudinal states. This leads to the resulting equations becoming linear with the measurement noise forming the process noise input. At times it turns out that even a parameter that strongly affects the airplane dynamics is estimated which vary widely among the approaches. This requires a careful look at the estimates. We also recommend a closer look at the correlation coefficients (that is generally ignored in such studies) of the estimated parameters which provide an insight into their subsequent uses. The present recipe has been shown to be better than the earlier approaches in estimating the unknowns. In particular the generalized cost functions that are introduced in the present work help to identify definitive results from deceptive results.  


\section{Introduction} 
In the previous paper an extensive study was carried out using an adaptive Extended Kalman Filter(EKF) tuning procedure called reference recursive recipe applied to the simulated data of a simple spring, mass, and damper system with a weak nonlinear spring. It is useful to further demonstrate the effectiveness of the above approach to handle more involved real data studies. It turns out that a good choice would be the flight test data analysis of an airplane system dealing with longitudinal and lateral motion. These generally have many states and measurements and a large number of aerodynamic parameters to be estimated. Further the flight tests cannot always be conducted in an ideal situation of the process noise and the measurement noises being white and Gaussian as is generally needed in the Kalman filter. The measurements are not available with respect to the center of gravity, possess scale and bias factors which will have to be modeled and estimated as well. The coupling between the longitudinal and lateral motion brings in added difficulty but makes the problem more interesting. At times the noisy measurements from the longitudinal and lateral motion are input into the longitudinal states. This leads to the resulting equations becoming linear with the measurement noise forming the process noise input. This is another example of subjectivity in estimation theory. However the final results should be meaningful, reasonable, acceptable and useful no matter whatever subjective inputs are introduced into the problem formulation and solution. 

\section{Aircraft Equations of Motion} 
The equations of motion for aircraft flight test data analysis are usually based on the following assumptions. The flat non rotating earth forms an inertial system, and with constant gravitational acceleration, the atmosphere is fixed relative to earth axes, and the aircraft is rigid with constant mass and moment of inertia, and so the effects of fuel sloshing, structural deformations, and the relative motion of control surfaces are negligible, further possess a vertical plane of symmetry, and the thrust is directed along the longitudinal body axis and through the aircraft center of gravity. 

\subsection{Aerodynamic Modeling} 
The most difficult part of aircraft dynamical analysis is the specification of the aerodynamic forces and moments acting on it and consequently most effort in aerospace studies involving analytical, computational, wind tunnel and flight test are directed towards it with the last being most realistic in terms of aircraft geometry, size and flight conditions.  

The aerodynamic forces and moments depend on the complete history of the aircraft motion involving airspeed, angles of incidence of the air with respect to the aircraft body, its linear and angular accelerations, control surface deflections, so a lot of effort goes in determining an adequate and useful aerodynamic model characterization. The aerodynamic modeling is a balance between complexity and acceptability for the purpose of control design, simulation, and flying qualities.   

\subsection{Linearization for System Identification}  

Data used for system identification applied to aircraft arise mostly from maneuvers that excite the longitudinal short period dynamics or the lateral modes associated with body axis roll, spiral motion, and lateral oscillations. In general the flight test data from the longitudinal and lateral motions should be decoupled but in practice there could be unavoidable coupling due to the airspeed changes and variations in the lateral quantities during a longitudinal maneuver, and vice versa. 

For flight data analysis the above coupling can be avoided in two ways. In the first linearization is carried out for small motions about a reference condition such as steady, wings level flight with no sideslip whence the equations decouple into the longitudinal motion in the plane of symmetry and the other the lateral motion out of the plane of symmetry. The resulting linearized models are adequate for (i) most flight conditions involving changes in linear and angular velocities from a reference condition, (ii) the aerodynamic effects described by linear functions of state and control variables, and (iii) the consequent stability analysis and control system design. 

The second method of handling the unavoidable coupling of longitudinal and lateral motion is to substitute the measured values in the nonlinear equations in order to linearize them. The resulting linear equations are fewer and with less number of model parameters. Here the dynamic pressure $\bar{q}$, airspeed V, and other variables that cause coupling are substituted by the measured values. For flight conditions near stall, spin, and maneuvers involving large angles and angular rates nonlinear models must be used. 

\subsection{Some Approaches in an Airplane Flight Test Data Analysis} 

For airplane flight test data analysis many approaches have been suggested to determine the parameters and the noise covariances. These are the natural formulation of Schultz (1976),the innovation formulation of Stepner and Mehra (1973), a combined formulation called MMLE3 of Maine and Iliff (1981), the combined formulation of Ishimoto (1997) to solve the practical problems in natural and innovation formulations. A more detailed discussion on aircraft flight test data analysis can be found in Klein and Morelli (2006) and Jategaonkar (2006). 

\subsubsection{Natural Formulation}
In the natural formulation the innovation cost \textbf{J} is directly minimized using a modified NR technique with respect to all the unknowns $\mathbf{P_0}$, $\Theta$, \textbf{Q}, and \textbf{R}. This method requires more computer time and memory than an output error method due to \textbf{Q} compulsively bringing in the Kalman filter. The computation of all the gradients dependent on the state and measurement functions and \textbf{Q} needs more computation. The serious convergence problems in this formulation to estimate $\Theta$ have been illustrated by Maine and Iliff (1981) for a simple scalar problem. 

\subsubsection{Innovation Formulation}
The unknown \textbf{Q} and \textbf{R} appear in \textbf{J} indirectly through innovation covariance and \textbf{K}. The innovation formulation trades \textbf{Q} and \textbf{R} for innovation covariance $\mathscr{R}=HPH^T+R$ and \textbf{K} which eliminates the convergence and computational problems associated with the estimation of \textbf{Q}. This estimation is done as a two step procedure. First an iteration of the NR algorithm is done to revise the estimates of all the parameters except for the term $HPH^T$. In the second the estimate of $HPH^T$ is revised, using the residuals form the previous step and the above steps are repeated until convergence. The elements of \textbf{K} are estimated along with the unknowns using the NR algorithm. Being simple is most widely used to solve the maximum likelihood estimation problem when both \textbf{R} and \textbf{Q} are unknown.

\subsubsection{Mixed Formulation}

This is a combined formulation, with the best features of the natural and innovation formulations in minimizing \textbf{J}. Instead of $\mathbf{P_0}$, \textbf{Q}, and \textbf{R} one works with innovation covariance $\mathscr{R}$, \textbf{K}, and \textbf{R} but should ensure a consistent \textbf{K} which can be in general be not square like innovation covariance and \textbf{R}.

\subsubsection{Combined Formulation}
This solves the practical problems of natural and innovation formulations and different from MMLE3. The unknown parameters are assumed to be included in the system transition matrices as well as \textbf{Q} and \textbf{R}. The filter provided innovation covariance $\mathscr{R}_f$ is compared with the ones computed from the residuals $\mathscr{R}_r$. In this method \textbf{J} is minimized subject to C($\Theta$) = diag($\mathscr{R}_f - \mathscr{R}_r$)=0. The purpose of this constraint is to accelerate the estimation of \textbf{R}. Although the sample covariance $\mathscr{R}_f$ is a function of the unknown parameters it is fixed during each optimization step and it is updated after every iteration. The Sequential Quadratic Programming method used to solve the above optimization problem provided substantial improvement in convergence.

One may ask the question as to why there are so many formulations for solving an optimization problem. The  reason is the unknowns do not occur in a simple way in the cost function, and there are many transformed variables with which one tries to solve for the basic unknowns. Further the size and the required compatibility conditions among the transformed variables lead to the many difficulties not found in the classical optimization problems.

\section{Analysis of Real Flight Test Case - 1}
\begin{figure}[h]
\includegraphics[width=6in,height=1.5in]{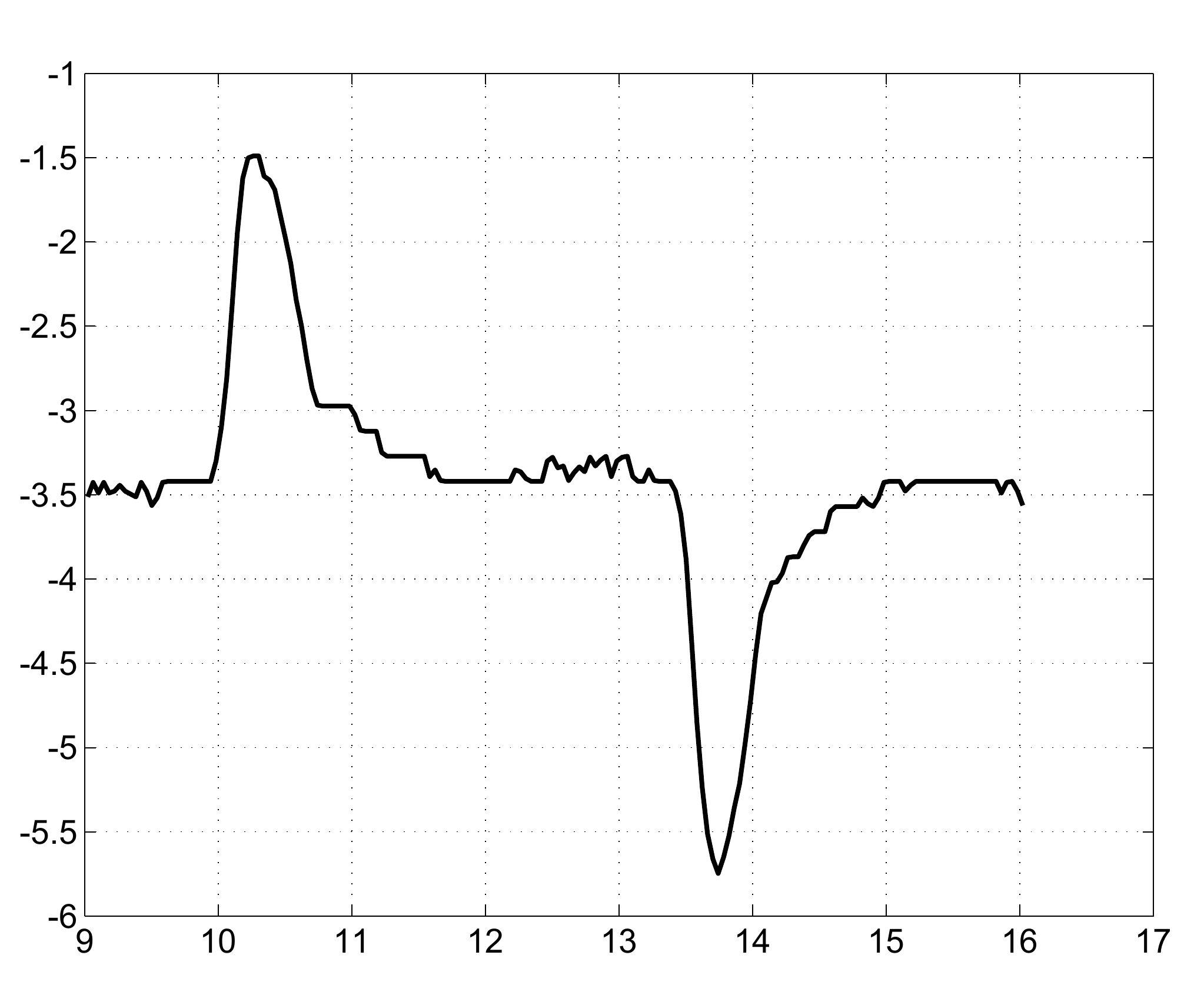}
\caption{Control input ($\delta_e$ in degrees) versus time (s)}
\label{input2}
\end{figure}

\begin{figure}[h]
\includegraphics[width=6in,height=1.5in]{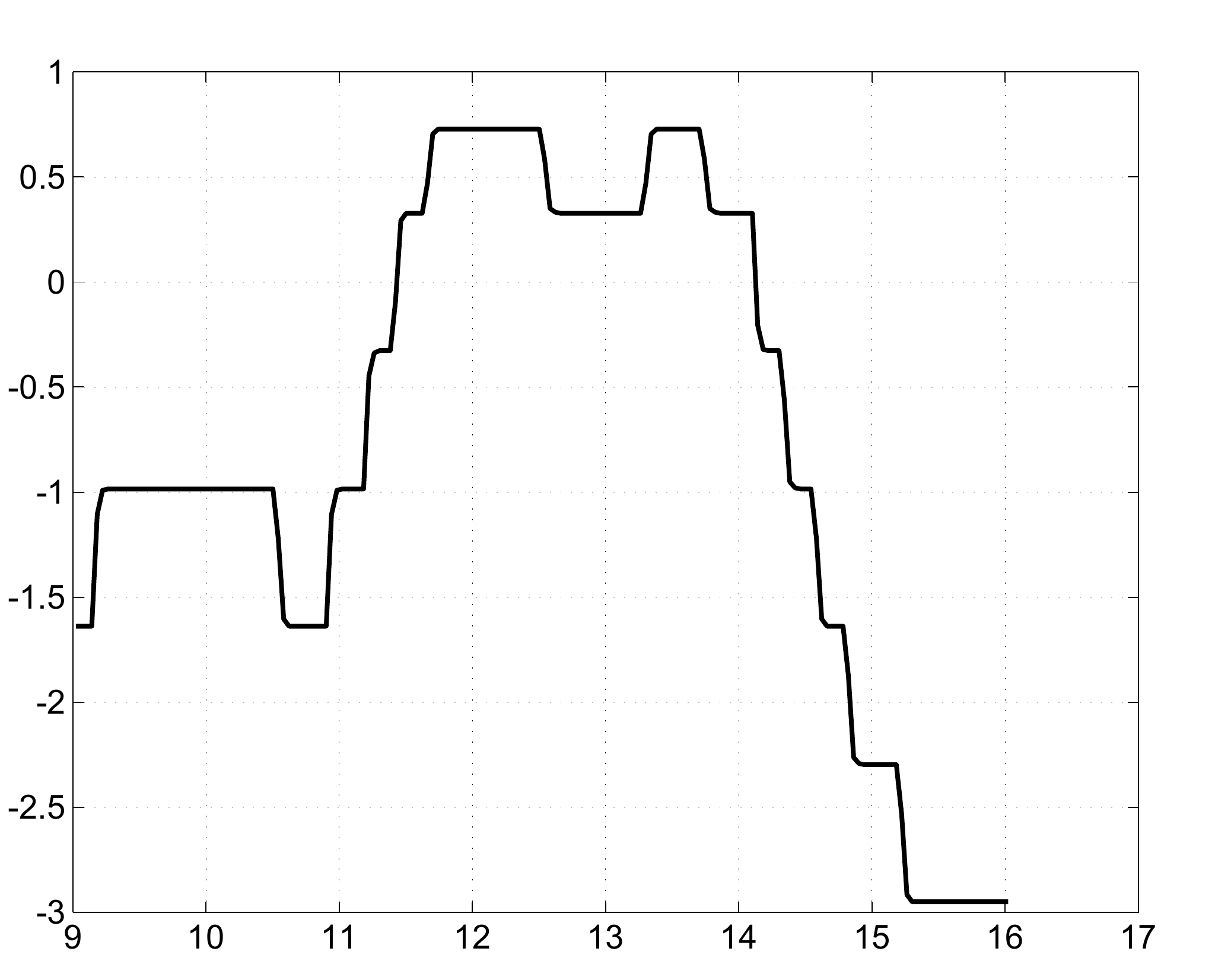}
\caption{Measurement input ($\phi_m$ in degrees) versus time (s)}
\label{case2_phi}
\end{figure}

\begin{figure}[h]
\includegraphics[width=6in,height=1.5in]{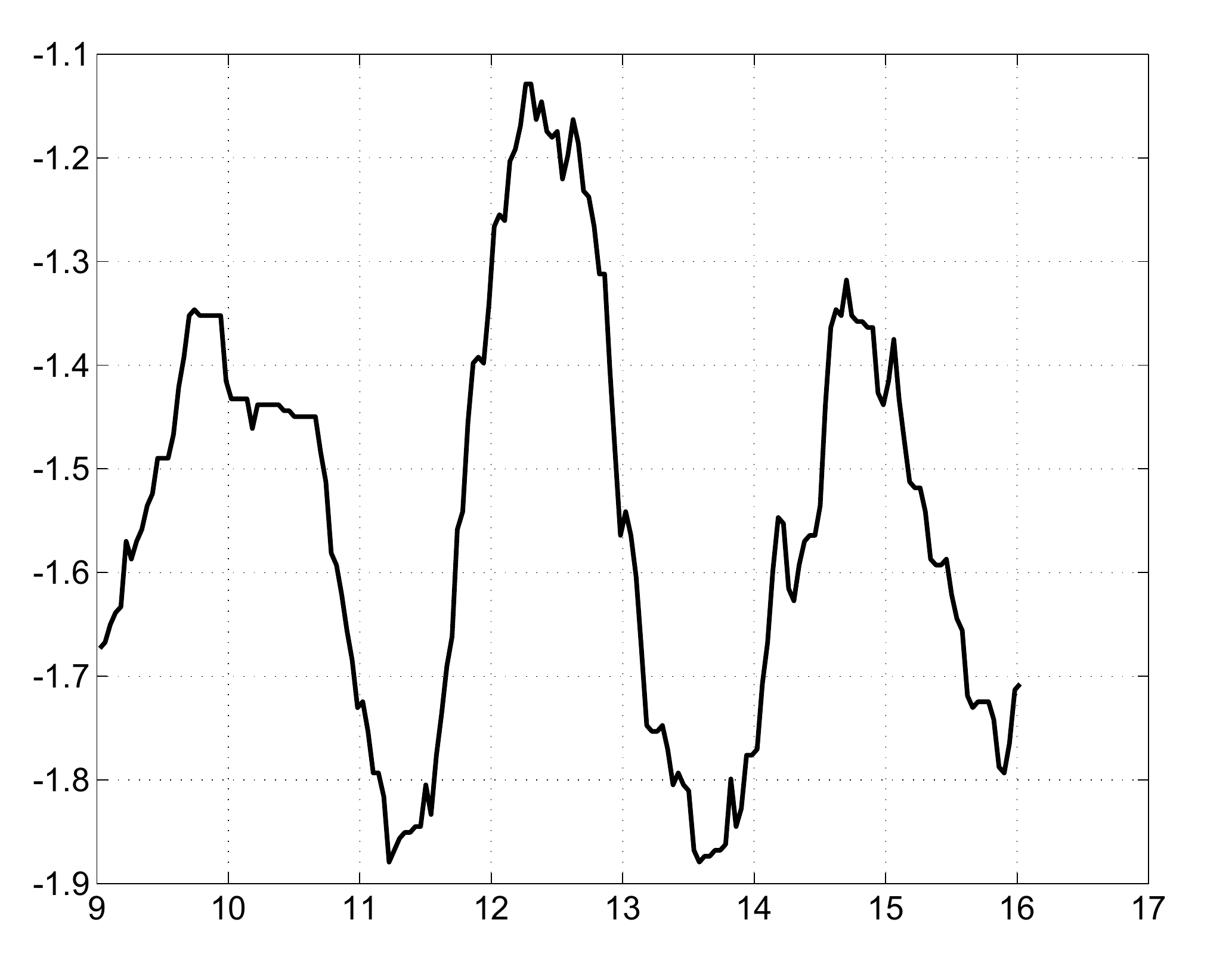}
\caption{Measurement input ($\beta_m$ in degrees) versus time (s)}
\label{case2_beta}
\end{figure}

\begin{figure}[h]
\includegraphics[width=6in,height=1.5in]{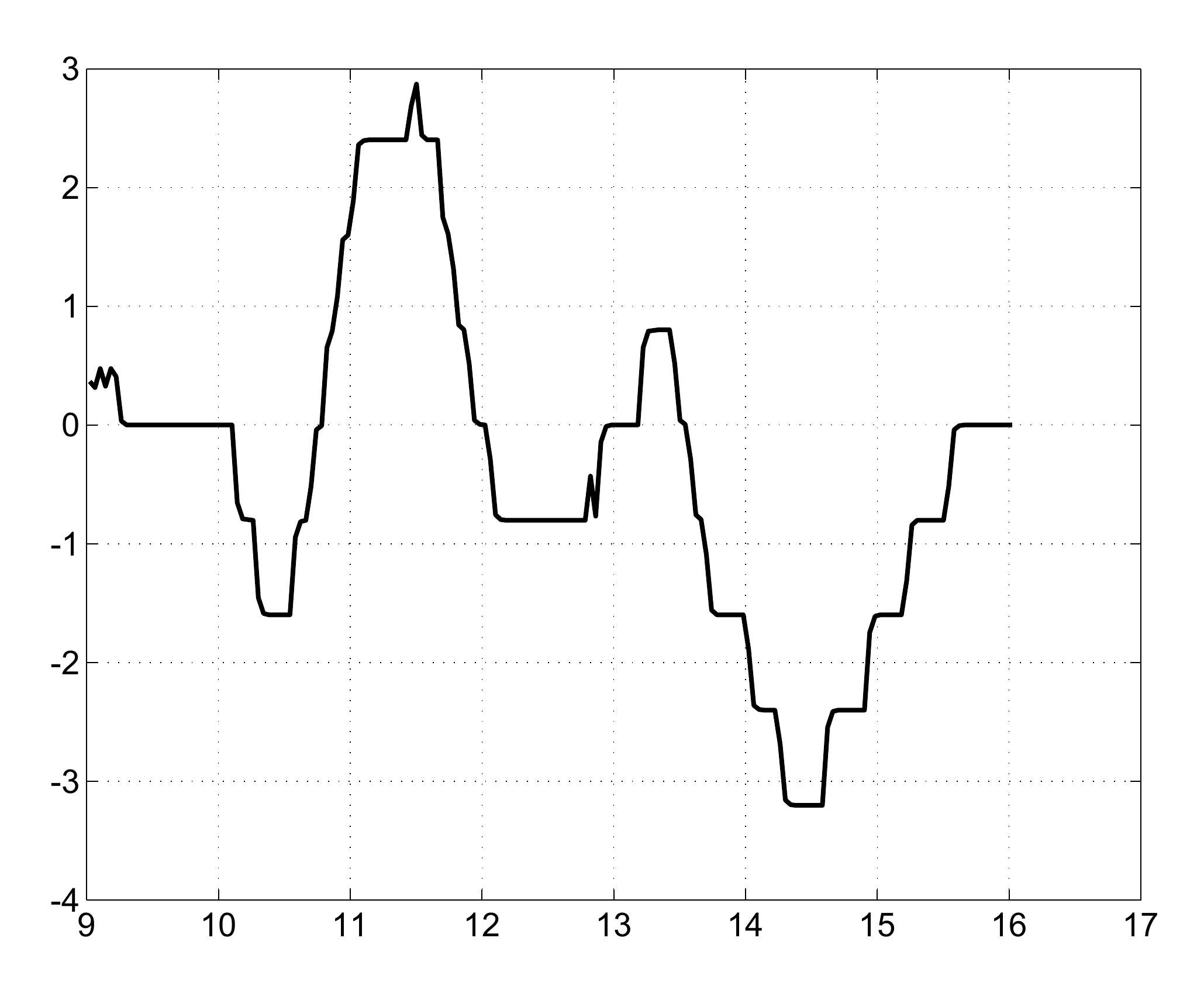}
\caption{Measurement input ($p_m$ in degrees) versus time (s)}
\label{case2_p}
\end{figure}

\begin{figure}[h]
\includegraphics[width=6in,height=1.5in]{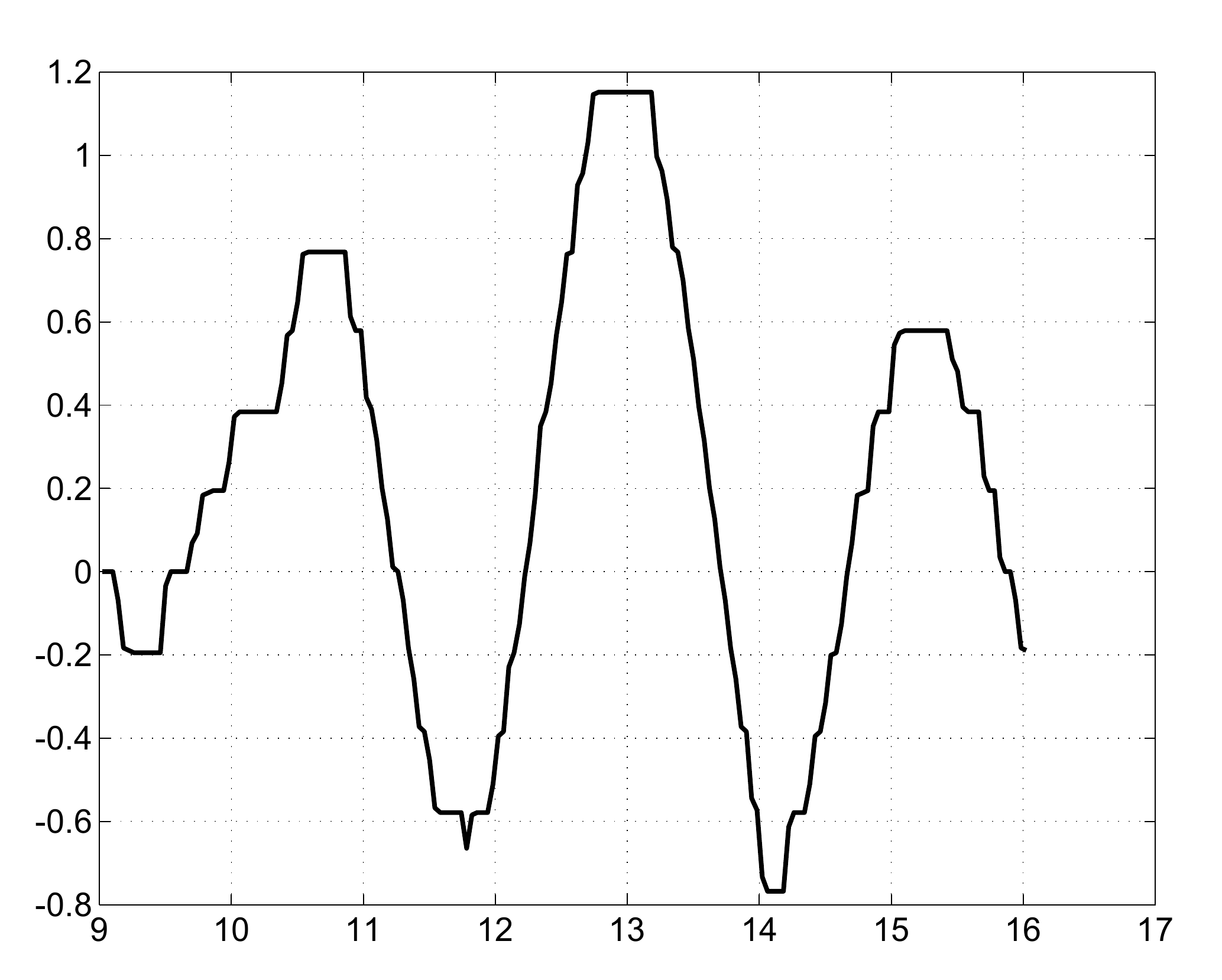}
\caption{Measurement input ($r_m$ in degrees) versus time (s)}
\label{case2_r}
\end{figure}

\begin{figure}[h]
\includegraphics[width=6in,height=3.5in]{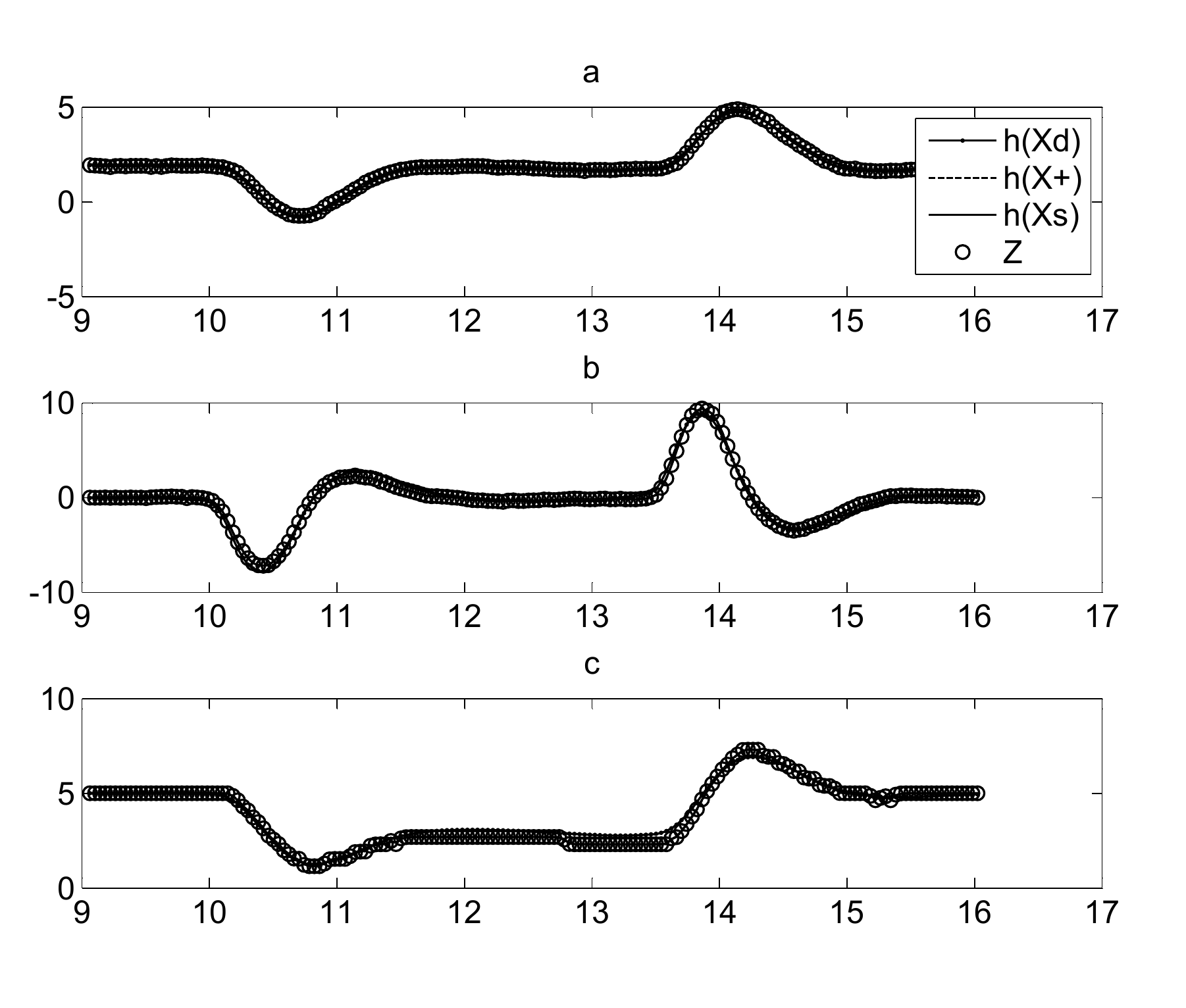}
\caption{Comparison of the predicted dynamics, posterior, smoothed and}
\caption*{measurement in degrees (a. angle of attack b. pitch rate c. pitch angle) vs time.}
\label{realQ2b1_s1}
\end{figure}

The salient features of this aircraft are available in NASA TM-X 56036 (Shafer 1975) and in NASA TP 1690  (Maine and Iliff 1981). The parameters are estimated in dimensionless form.  The data set obtained is for a short period motion excited by the up and down elevator control input ($\delta_e$ in degrees) as shown in Fig.\ref{input2}. In general the flight test data is such that the longitudinal and lateral motions are decoupled. Some of the available measurements have been used as inputs in the state equations which includes roll angle ($\phi_m$), sideslip ($\beta_m$), roll rate ($p_m$), yaw rate ($r_m$) and the angle of attack ($\alpha_m$) are shown in Fig.\ref{case2_phi}, Fig.\ref{case2_beta}, Fig.\ref{case2_p}, Fig.\ref{case2_r} and Fig.\ref{realQ2b1_s1} respectively. The state equations ($n=3$) for the angle of attack ($\alpha$), pitch rate (q) and the pitch angle ($\Theta$) respectively are 
\begin{align*}
\dot{\alpha}=&-\frac{\bar qS}{mV}C_L+q+\frac{g}{V} (cos(\phi_m)cos(\alpha_m)cos(\theta)+sin(\alpha_m)sin(\theta))- \\
&\beta_m(p_mcos(\alpha_m)+r_msin(\alpha_m))\\
\dot{q}=&\frac{\bar q S \bar c}{Iyy}(C_{m_\alpha}\alpha+C_{m_q}\frac{\bar c}{2V}q+C_{m_{\delta_e}}\delta_e+C_{m_0})+
\frac{Izz-Ixx}{Iyy}r_mp_m\\
\dot{\theta}=&qcos(\phi_m)-r_msin(\phi_m)+\theta_0
\end{align*}
The measurement equations ($m$=5) are
\begin{align*}
{\alpha_m}&=\alpha-K_\alpha x_\alpha \frac{q}{V} \\
{q_m}&=q\\
{\theta_m}&=\theta\\
{a_{n_m}}&=\frac{\bar q S}{mg}C_N+\frac{x_{a_n}}{g}\dot{q} \\
{a_{x_m}}&=-\frac{\bar q S}{mg}C_A+\frac{z_{a_x}}{g}\dot{q}
\end{align*}
where
\begin{align*}
C_L=&C_Ncos(\alpha)-C_Asin(\alpha)+C_{L_0}\\
C_N=&C_{N_\alpha}\alpha+C_{N_{\delta_e}}\delta_e+C_{N_0}\\
C_A=&C_{A_\alpha}\alpha+C_{A_{\alpha^2}}\alpha^2+C_{A_{\delta_e}}\delta_e+C_{A_0}
\end{align*}


The unknown parameter set ($p=13$) is $\Theta=(C_{N_{\alpha}},C_{N_{\delta_e}},C_{L_{0}},C_{m_{\alpha}},C_{m_{q}},C_{m_{\delta_e}},C_{m_{0}},\theta_0,C_{N_0},C_{A_{\alpha}},\\C_{A_{\alpha^2}},C_{A_{\delta_e}},C_{A_0})^T$. The ones with suffix `$\delta_e$' are the control derivatives, the ones with suffix zero are the biases and all others are aerodynamic derivatives. The initial states are taken as initial measurement and the initial parameter values are taken as $(4,0.24,0.17,-0.48,-17,-0.9,-0.05,-0.02,\\0.175,-0.3,0.03,-0.083,-0.015)^T$. 

\begin{table}[h]
\begin{center}
\caption*{Other constant values used for case-1}{}
\begin{tabular}{| c | c | c | c | c | c | }
\hline
$\bar c$=5.58 & S=184 & m=172.667 & Ixx=4142.9  &  Iyy=3922.4 & Izz=7642.5  \\ \hline
g=32.2 & V=403.1 & $\bar q$=83.08 & $K_\alpha x_\alpha$=-0.0279 & $x_{a_n}$=0.101 & $z_{a_x}$=-1.17\\ \hline
\end{tabular}
\end{center}
\end{table}

In the analysis of real flight data the following filter outputs are studied which provide an insight into the filter performance. The convergence of the following quantities through the iterations are analyzed. 

\begin{enumerate}
\item The parameter estimates $\Theta$ and their covariances $P_{\Theta}$.
\item The noise covariance of \textbf{Q} and \textbf{R}. 
\item The state dynamics without measurement and process noises based on the estimated parameter after the filter pass through the data Xd, the prior state X-, the posterior state X+, the smoothed state Xs and the measurement Z.
\item The sample innovation, filtered residue and the smoothed residue along with their bounds which is the square root of the predicted covariances given respectively by ($\textbf{R}+H_kP_{k|k-1}H_k^T$), ($\textbf{R}-H_{k|k}P_{k|k}H_{k|k}^T$) and ($\textbf{R}-H_{k|N}P_{k|N}H_{k|N}^T$) by the filter.
\item The estimated measurement and process noise samples as well as their autocorrelations.
\item The cost functions (\textbf{J1-J8}) after the final convergence.
\end{enumerate} 

\begin{figure}[h]
\includegraphics[width=6in,height=3in]{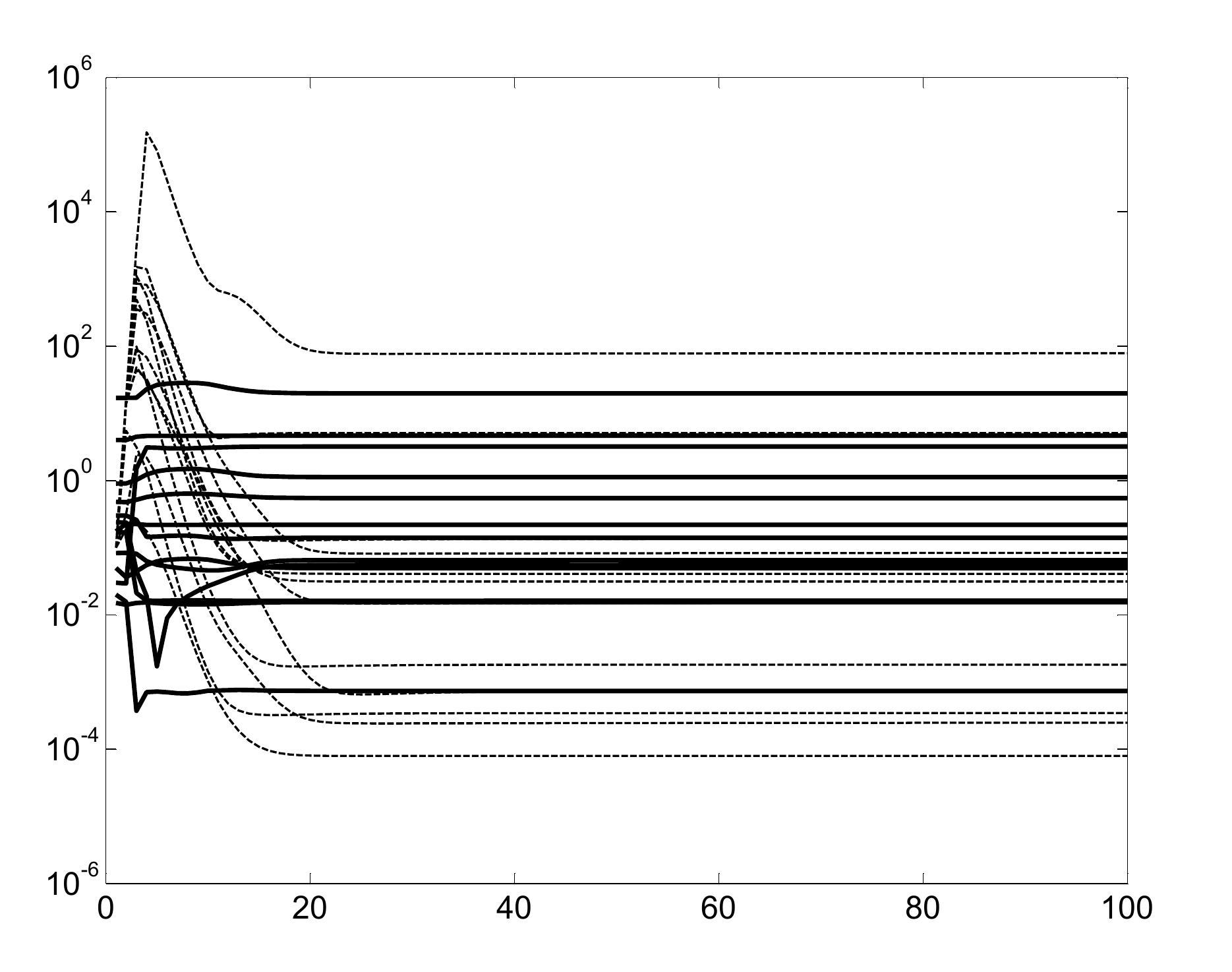}
\caption{Variation of initial parameters $\Theta_0$(continuous) and its $\mathbf{P_0}$(dashed) with iterations}
\label{realQ2b1_P0}
\end{figure}

\begin{figure}[h]
\includegraphics[width=6in,height=2.5in]{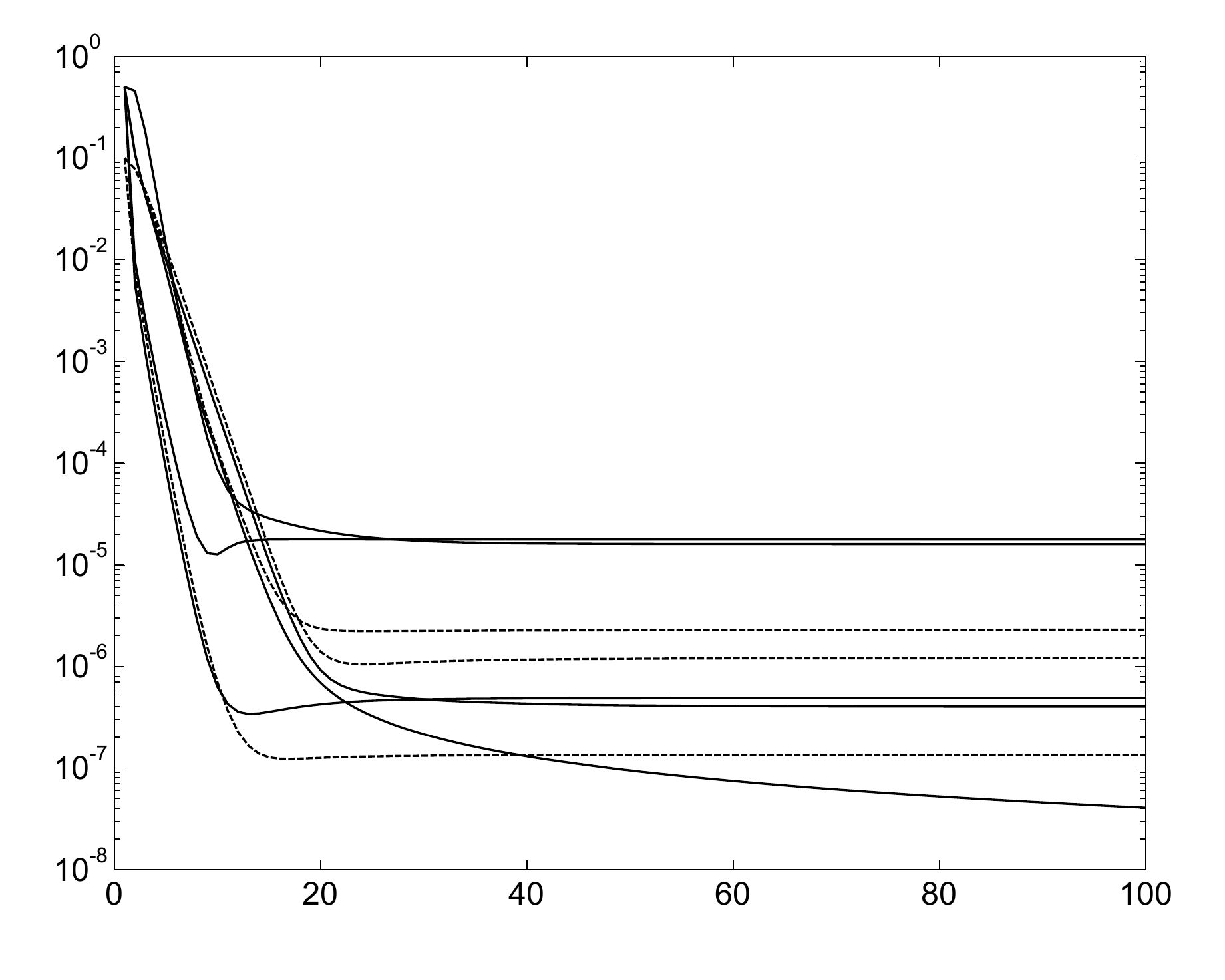}
\caption{Variation of \textbf{Q} (dashed) and \textbf{R} (continuous) with iterations}
\label{realQ2b1_R}
\end{figure}

\begin{figure}[h]
\includegraphics[width=6in,height=2.5in]{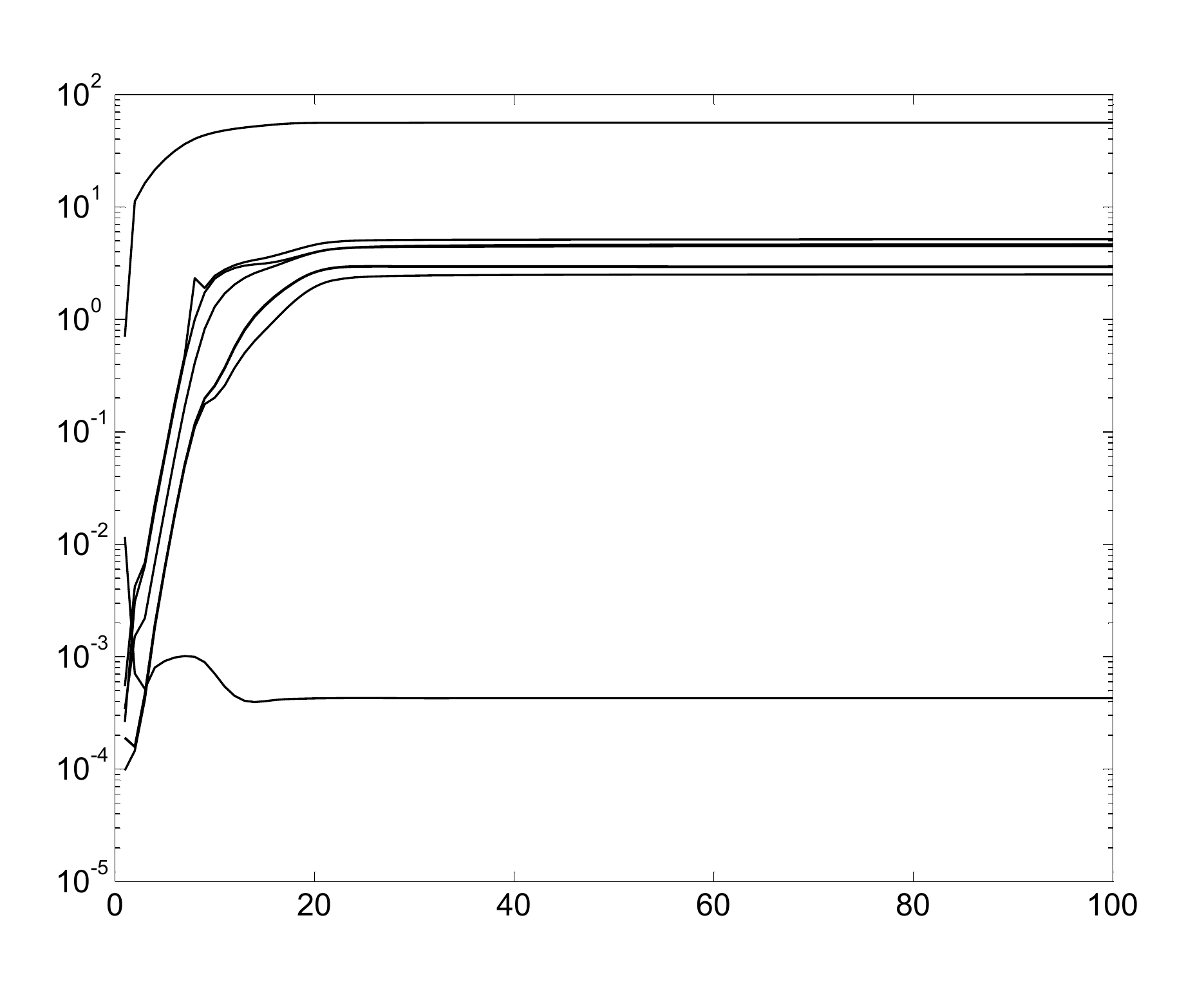}
\caption{Variation of different costs (\textbf{J1-J8}) with iterations}
\label{realQ2b1_J}
\end{figure}

\begin{figure}[h]
\includegraphics[width=6in,height=2.5in]{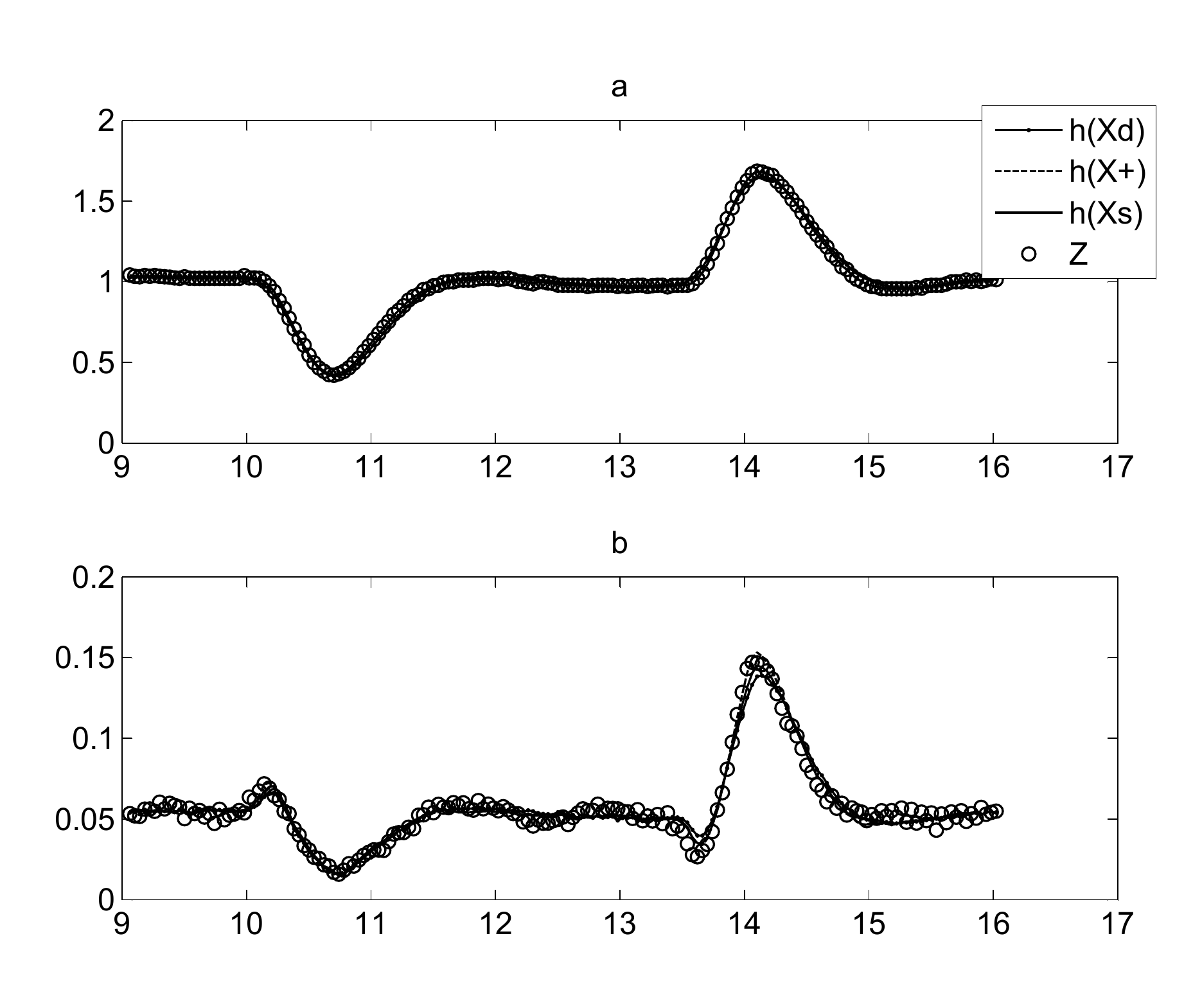}
\caption{Comparison of the predicted dynamics, posterior, smoothed and}
\caption*{measurement in $ft/sec^2$ (a.normal acceleration b.axial acceleration) vs time}
\label{realQ2b1_h4}
\end{figure}

\begin{figure}[h]
\includegraphics[width=6in,height=2in]{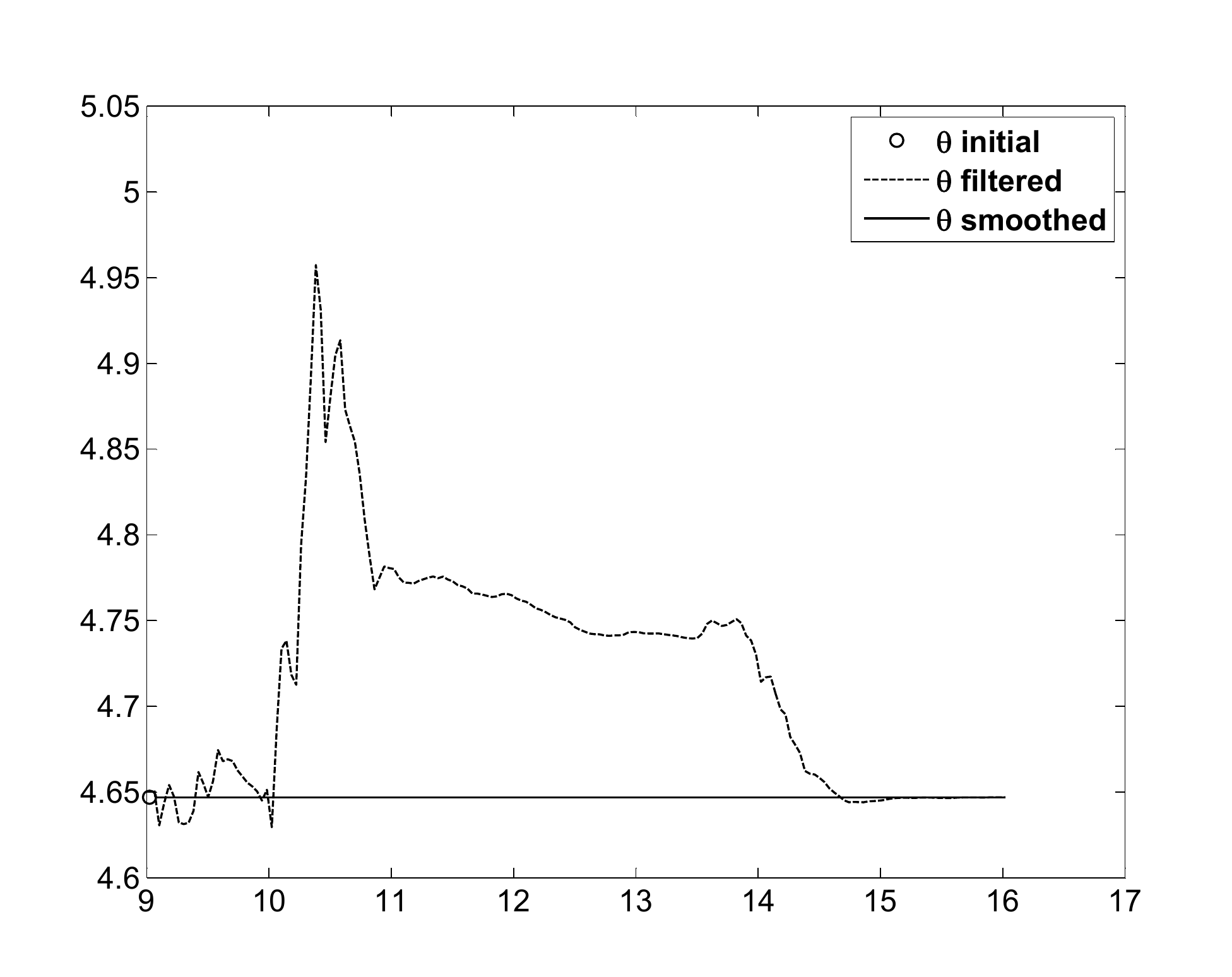}
\caption{Variation of the parameter estimate ($C_{N_\alpha}$) through time instants}
\label{CNalpha}
\end{figure}

Case-1 real data is run using the reference EKF (\textbf{Q} $>$ 0) with 100 iterations. It turns out that the off diagonal elements of the correlation coefficient matrix with Q = 0 reduced substantially when Q $>$ 0 in all the present case studies thus indicating the latter estimates more trustworthy. The Fig.\ref{input2}-\ref{case2_r} show the inputs used in state equations. The Fig.\ref{realQ2b1_P0} shows the variation of parameter estimates and its initial covariance $\mathbf{P_0}$ with iterations and a similar Fig.\ref{realQ2b1_R} for \textbf{Q} and \textbf{R}. The values of \textbf{J1-J3} are close to the number of measurements ($m=5$) with \textbf{J6-J8} are close to the number of states ($n=3$) as shown in Fig.\ref{realQ2b1_J} and Table-\ref{tbcase2Q}. This means the measurement and state equations are well balanced. The \textbf{J5} is the negative log likelihood cost function. The later Fig.\ref{realQ2b1_s1}-\ref{realQ2b1_h4} compares (i) the state dynamics based on the estimated parameter after the filter pass through the data, (ii) the state after measurement update, (iii) the smoothed state and (iv) the measurement.  Unlike in the simulated studies, the estimated measurement and process noise did not have constant statistical characteristics across time. Another experiment was carried out by generating a typical data set by using the estimated theta and injecting the estimated \textbf{Q} and \textbf{R} as additive white Gaussian noise. This is to determine the effect of non White and non Gaussian noise distribution in the real data on the CRBs. After each iteration in the reference recipe the $\Theta$, \textbf{Q} and \textbf{R} were reset as from the real data. Similar experiment was also conducted by updating $\Theta$ as well. It turned out that there is not much of a difference in the final estimates and the CRBs. Two other filter runs were carried out using the MT and MS statistics for the estimation of \textbf{Q} and \textbf{R} with scaled up $\mathbf{P_0}$. The behaviour of the various cost function and in particular \textbf{J6} and \textbf{J7} in Table-\ref{tbcase2QMTMS} shows that the choice of the filter statistics for estimating \textbf{Q} and \textbf{R} in the proposed reference approach is the best possible when compared to other approaches.  Another feature of recursive parameter estimation is that it can vary through time instants and point to two distinct values as reflected in the estimation of $C_{N_{\alpha}}$ in Fig.\ref{CNalpha}. This feature of tracking time varying parameters by the EKF brings in clearly another advantage of sequential processing instead of batch processing of the data by a least squares (LS) procedure. If LS had been used then for the parameter `$C_{N_\alpha}$' only an average value would have been obtained. If one persists in using a batch processing procedure then to get the varying parameter such a feature should be modelled which would include where to change and one might have to use the data in blocks by splitting them and all such exercises have to be carried out which are not easy.

\subsection{Remarks on Case - 1}   
\begin{figure}[h]
\includegraphics[width=6in,height=2.5in]{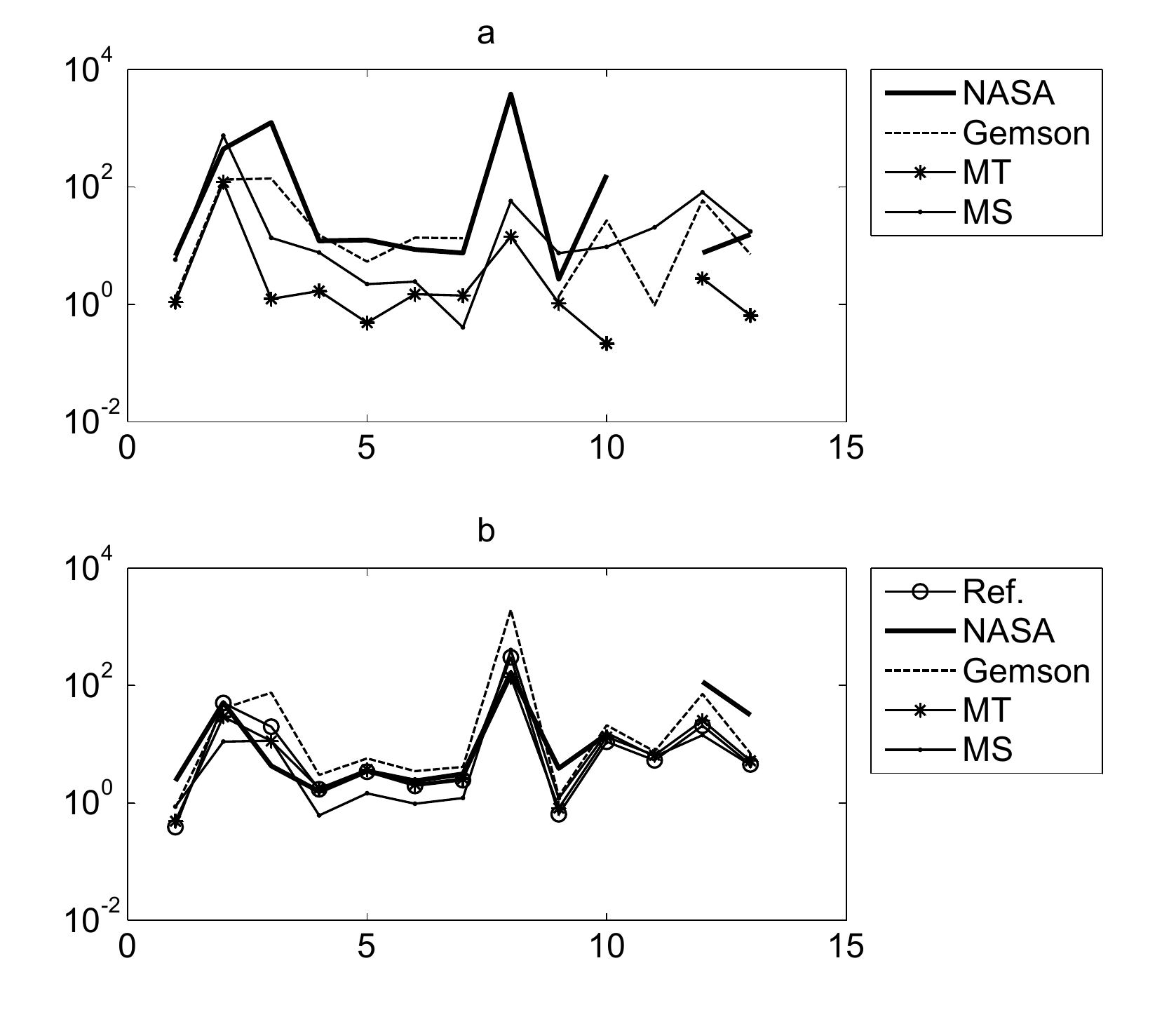}
\caption{Comparison of (a) Absolute percentage error with respect to the reference of the parameter estimates and (b) $\%$CRBs  by different methods}
\label{comp2}
\end{figure}

The NASA results have been generated assuming \textbf{Q} = 0 and are comparable with reference procedure for the parameter estimates and their CRBs. Further the MT and MS methods give quite different estimates for the \textbf{Q} and \textbf{R} values than from the reference case. We believe that the reference procedure provides the best possible parameter estimates and their uncertainties. The percentage CRB of the $i^{th}$ parameter estimate is defined as \%CRB=$100\times\sqrt{P_{\Theta_{ii}}}/|\Theta_i|$. From the plot of \% error in the parameter estimate with respect to the reference value and the \%CRB in Fig \ref{comp2}, it can be seen that the parameters $\theta_0$ and $C_{A_{\delta_e}}$ are relatively weak when compared to other parameters. The CRBs as estimated by different methods generally appear to vary widely. However what is interesting is that even the estimate of the strong parameter $C_{m_{q}}$ varies widely among the methods. Such a behaviour of the filter across the parameter estimates shows how important is the tuning of the filter statistics namely $\mathbf{P_0}$, \textbf{Q} and \textbf{R} in parameter estimation and their uncertainties. The rounded off 100$\times$C matrix of the parameter estimates for this case is 
\begin{center}
\begin{tabular}{|c|c|c|c|c|c|c|c|c|c|c|c|c|c|}
\hline

$\Theta$ & $C_{N_{\alpha}}$ & $C_{N_{\delta_e}}$ & $C_{L_{0}}$ & $C_{m_{\alpha}}$ & $C_{m_{q}}$ & $C_{m_{\delta_e}}$ & $C_{m_{0}}$ & $\theta_0$ & $C_{N_0}$ & $C_{A_{\alpha}}$ & $C_{A_{\alpha^2}}$ & $C_{A_{\delta_e}}$ & $C_{A_0}$\\ \hline
$C_{N_{\alpha}}$ & 100	&	62	&	0	&	-18	&	0	&	-7	&	-3	&	0	&	32	&	-1	&	-10	&	-10	&	-8	\\ \hline
$C_{N_{\delta_e}}$  & 62	&	100	&	0	&	-10	&	1	&	-11	&	-10	&	0	&	93	&	-4	&	-2	&	-16	&	-15	\\ \hline
 $C_{L_{0}}$ & 0	&	0	&	100	&	0	&	0	&	0	&	0	&	0	&	0	&	0	&	0	&	0	&	0	\\ \hline
$C_{m_{\alpha}}$ & -18	&	-10	&	0	&	100	&	18	&	51	&	31	&	0	&	-5	&	-15	&	1	&	-15	&	-7	\\ \hline
 $C_{m_{q}}$ &  0	&	1	&	0	&	18	&	100	&	76	&	79	&	0	&	1	&	6	&	-3	&	5	&	2	\\ \hline
 $C_{m_{\delta_e}}$ & -7	&	-11	&	0	&	51	&	76	&	100	&	97	&	0	&	-10	&	-1	&	-2	&	-14	&	-14	\\ \hline
$C_{m_{0}}$ & -3	&	-10	&	0	&	31	&	79	&	97	&	100	&	0	&	-11	&	2	&	-3	&	-11	&	-15	\\ \hline
  $\theta_0$ & 0	&	0	&	0	&	0	&	0	&	0	&	0	&	100	&	0	&	0	&	0	&	0	&	0	\\ \hline
 $C_{N_0}$ & 32	&	93	&	0	&	-5	&	1	&	-10	&	-11	&	0	&	100	&	-5	&	2	&	-15	&	-15	\\ \hline
 $C_{A_{\alpha}}$ & -1	&	-4	&	0	&	-15	&	6	&	-1	&	2	&	0	&	-5	&	100	&	-83	&	42	&	6	\\ \hline
 $C_{A_{\alpha^2}}$ & -10	&	-2	&	0	&	1	&	-3	&	-2	&	-3	&	0	&	2	&	-83	&	100	&	-7	&	16	\\ \hline
 $C_{A_{\delta_e}}$ & -10	&	-16	&	0	&	-15	&	5	&	-14	&	-11	&	0	&	-15	&	42	&	-7	&	100	&	91	\\ \hline
 $C_{A_0}$& -8	&	-15	&	0	&	-7	&	2	&	-14	&	-15	&	0	&	-15	&	6	&	16	&	91	&	100	\\ \hline
\end{tabular}
\end{center}

\begin{table}[h]
\caption{Real flight test data case-1 results ($\Theta, \sigma_\Theta$).}
\label{tbcase2Q}
\begin{center}
\begin{tabular}{|c|c| c| c| c| c| c| c| c|c|c|c|}
\hline
$\Theta$ &
\makecell{Reference} &   
\makecell{NASA} &   
\makecell{Gemson} &  
\makecell{MT} & 
\makecell{MS} 
\\ \hline

\makecell{$C_{N_{\alpha}}$ \\ \\ $C_{N_{\delta_e}}$ \\ \\ $C_{L_{0}}$ \\ \\ $C_{m_{\alpha}}$ \\ \\ $C_{m_{q}}$ \\ \\ $C_{m_{\delta_e}}$ \\ \\ $C_{m_{0}}$ \\ \\ $\theta_0$ \\ \\ $C_{N_0}$ \\ \\ $C_{A_{\alpha}}$ \\ \\ $C_{A_{\alpha^2}}$ \\ \\ $C_{A_{\delta_e}}$ \\ \\ $C_{A_0}$ \\} &

\makecell{  4.6469 \\ (0.0179) \\  0.0555\\  (0.0277)  \\  0.0162\\   (0.0032)  \\  -0.5468 \\  (0.0093) \\ -19.8027  \\  (0.6692) \\  -1.1229  \\  (0.0218) \\  -0.0495  \\   (0.0012) \\ 0.0007  \\ (0.0021) \\  0.2195 \\   (0.0014)  \\ -0.1398 \\ (0.0153) \\  -3.2088 \\  (0.1702) \\  -0.0651 \\   (0.0134)   \\ -0.0155 \\  (0.0007)} &

\makecell{ 4.9584 \\  (0.1168) \\ 0.3023 \\  (0.1550) \\  0.2189 \\   (0.009344) \\  -0.6125 \\  (0.00953) \\ -22.27 \\   (0.7713) \\  -1.2193 \\   (0.02881) \\  -0.0532\\   (0.00165) \\ 0.0273 \\ (0.04518) \\  0.2254 \\  (0.008725) \\ -0.3639 \\ (0.05328) \\ --\\ (--) \\  -0.07 \\  (0.08084) \\ -0.0131\\   (0.004088)} &

\makecell{ 4.7073 \\ (0.039) \\  0.1292 \\ (0.0523) \\  -0.0064 \\ (0.0048)  \\  -0.63  \\ (0.0188) \\-20.8623 \\  (1.1908) \\-1.2763 \\ (0.0442) \\ -0.0561 \\ (0.0023) \\ 0.0007  \\  (0.0135)  \\ 0.2225  \\ (0.0029)   \\ 0.0214 \\ (-0.1023) \\  -3.2397  \\ (0.2430) \\ -0.0267 \\ (0.0191) \\ -0.0144 \\(0.0010)} &

\makecell{  4.6978 \\  (0.0229)  \\ 0.1225  \\ (0.0357)  \\ 0.0160 \\  (0.0018) \\ -0.5560 \\ (0.0098) \\ -19.7062  \\ (0.7286)  \\ -1.1396 \\ (0.0236)  \\ -0.0502 \\(0.0013) \\ 0.0008 \\ (0.0011)  \\  0.2218  \\(0.0018)   \\ -0.1401 \\ (0.0185)  \\ -3.2088  \\(0.2070)  \\  -0.0633 \\ (0.0160) \\ -0.0154 \\ (0.0008)} &

\makecell{  4.9141 \\ (0.0422) \\  0.4691 \\  (0.0517)  \\  0.0184 \\  (0.0021)  \\ -0.5885 \\(0.0036) \\ -20.2395 \\ (0.2937)  \\ -1.1503 \\ (0.0111)  \\ -0.0497 \\(0.0006) \\ 0.0003  \\ (0.0012)  \\ 0.2358  \\ (0.0028)   \\-0.1265 \\ (0.0197) \\ -3.8625  \\ (0.2376)  \\-0.1178 \\ (0.0167)  \\ -0.0182 \\  (0.0008) } 
 
\\ \hline 

\end{tabular}
\end{center}
\vspace{0.5cm}
\caption{Real flight test data case-1 results* (\textbf{R,Q,J}).}
\label{tbcase2QMTMS}
\begin{center}
\begin{footnotesize}
\begin{tabular}{|c| c| c|| c| c| c|| c|c|c|}
\hline 

\makecell{\textbf{R} (Ref)\\ $\times10^{-6}$ }&  
\makecell{\textbf{Q} (Ref)\\ $\times10^{-6}$}&  
\makecell{\textbf{J1-J8} \\(Ref) }&

\makecell{\textbf{R} (MT)\\ $\times10^{-6}$ }&  
\makecell{\textbf{Q} (MT)\\ $\times10^{-6}$}&  
\makecell{\textbf{J1-J8} \\(MT) }&

\makecell{\textbf{R} (MS) \\ $\times10^{-6}$}&  
\makecell{\textbf{Q} (MS)\\ $\times10^{-6}$}&  
\makecell{\textbf{J1-J8} \\(MS) }
\\ \hline

\makecell{ 0.49  \\  0.04  \\  0.40  \\  15.98  \\  17.70} &
\makecell{ 0.134  \\  2.287  \\  1.204} &
\makecell{4.4752  \\  5.1532  \\  4.6432  \\  0.0004 \\ -56.2206  \\  2.9551 \\   2.9303 \\ 2.5161} &

\makecell{0.4107  \\  0.0312  \\  3.9381 \\  94.5086 \\  26.3511} &
\makecell{0.0393 \\   2.6418  \\  0.3231} &
\makecell{ 4.0090 \\   3.9630  \\  2.9764 \\   0.0004 \\ -54.7596 \\   6.6681 \\   6.4985 \\ 2.4562} &

\makecell{ 3.2046 \\  37.6770 \\   7.5509 \\ 198.2716 \\  28.9841} &
\makecell{0.0001  \\  0.0015  \\  0.3456} &
\makecell{  3.3893 \\   3.3866  \\  3.2057 \\   0.0002 \\ -49.6223  \\  3.8921   \\ 4.7110 \\ 2.6369} 

\\ \hline

\end{tabular}
\end{footnotesize}
\\ *Cost functions are not close to their expected values in MT and MS methods.
\end{center}
\end{table}


\clearpage
\section{Analysis of Real Flight Test Case - 2}
\par
\begin{figure}[h]
\includegraphics[width=6in,height=2in]{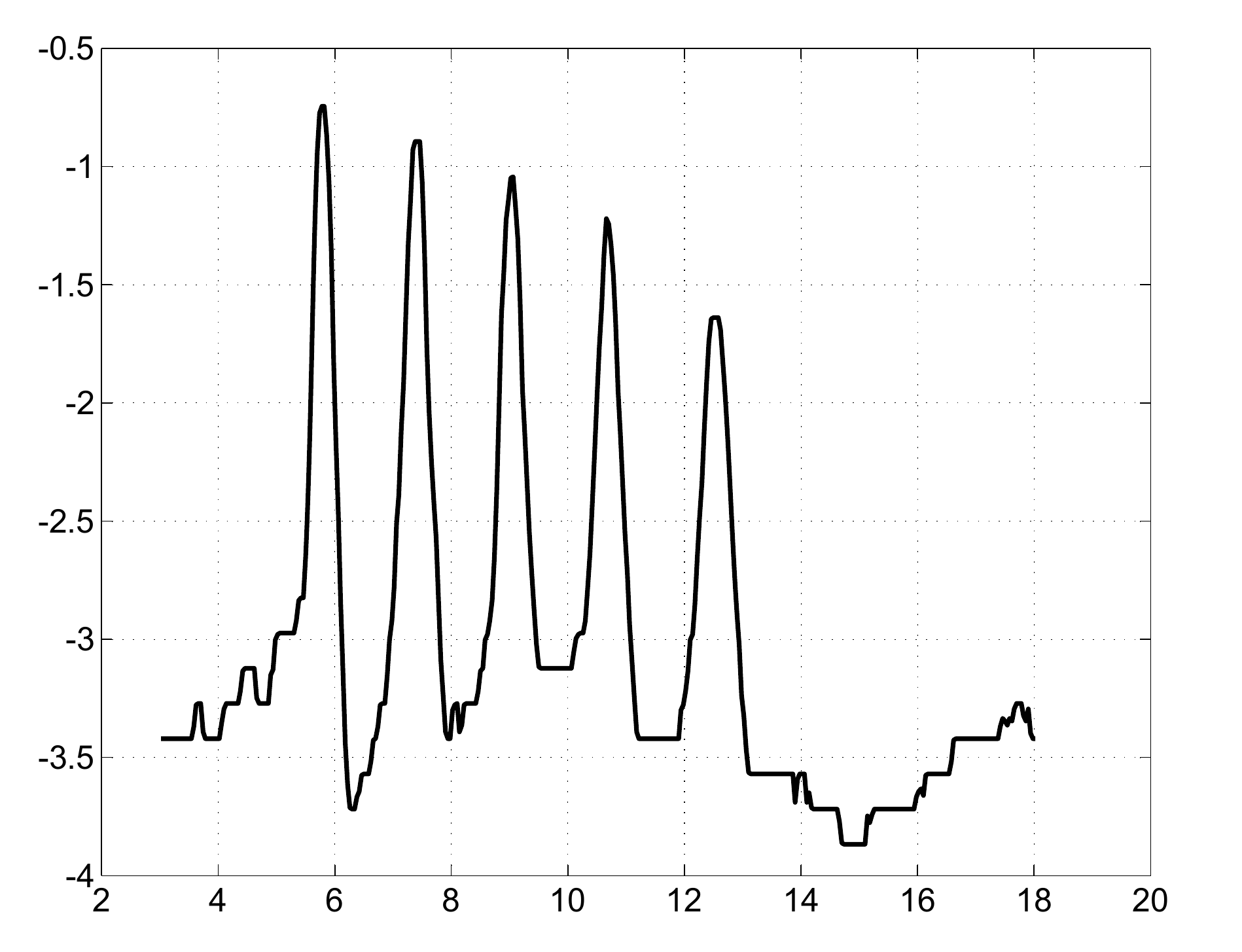}
\caption{Control input ($\delta_e$ in degrees) versus time (s)}
\label{input3}
\end{figure}

\begin{figure}[h]
\includegraphics[width=6in,height=2in]{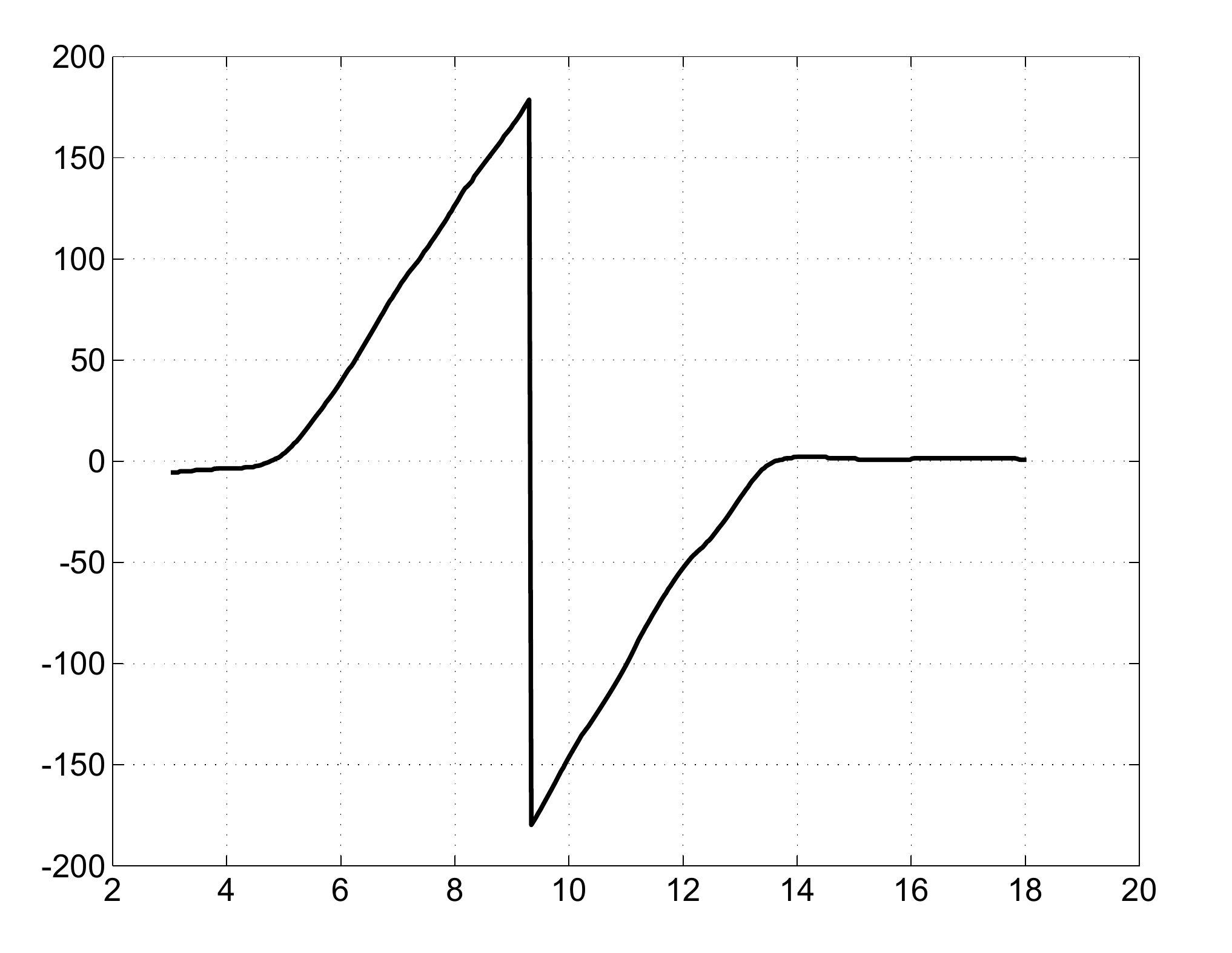}
\caption{Measurement input ($\phi_m$ in degrees) versus time (s)}
\label{case3_phi}
\end{figure}

\begin{figure}[h]
\includegraphics[width=6in,height=2in]{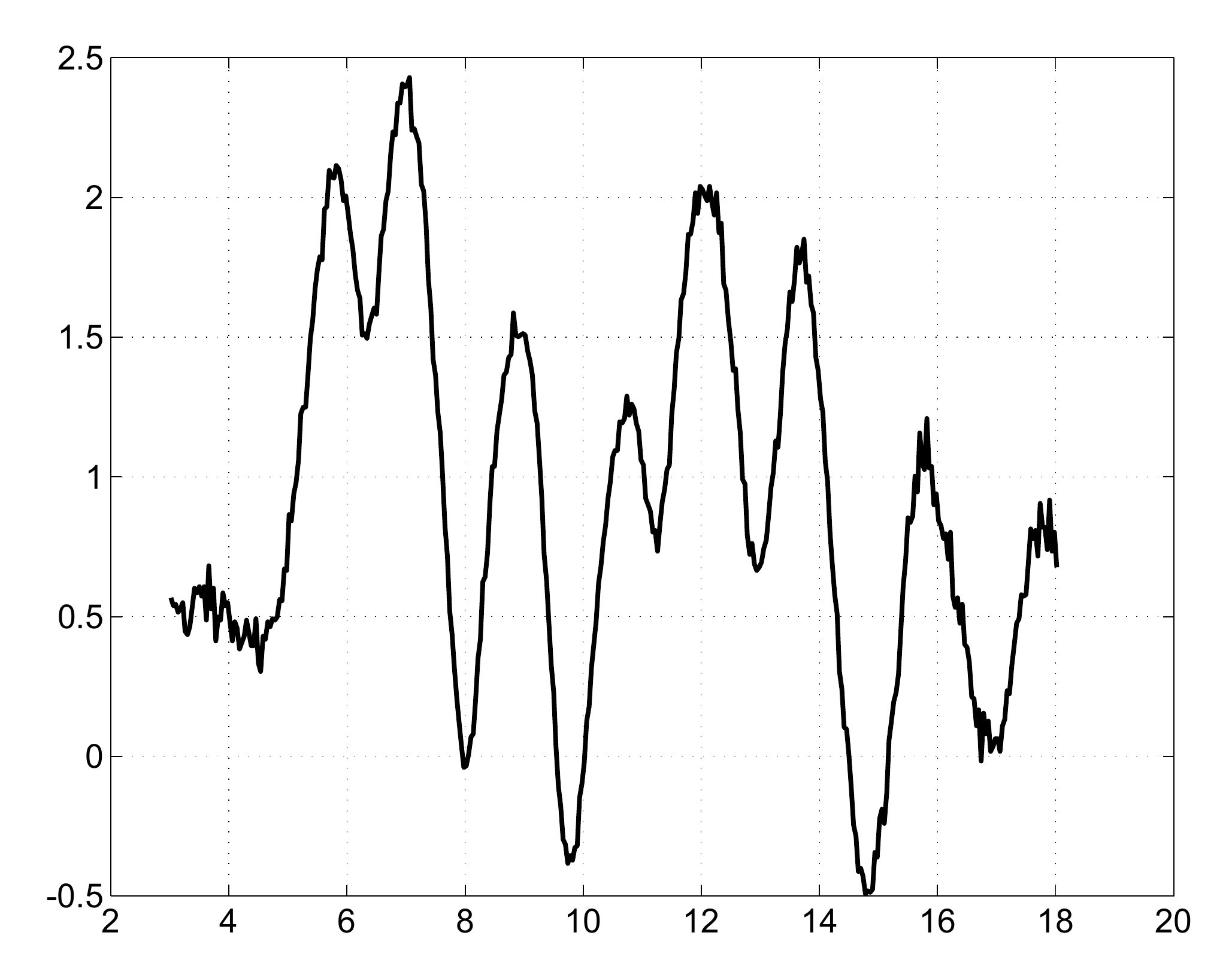}
\caption{Measurement input ($\beta_m$ in degrees) versus time (s)}
\label{case3_beta}
\end{figure}

\begin{figure}[h]
\includegraphics[width=6in,height=2in]{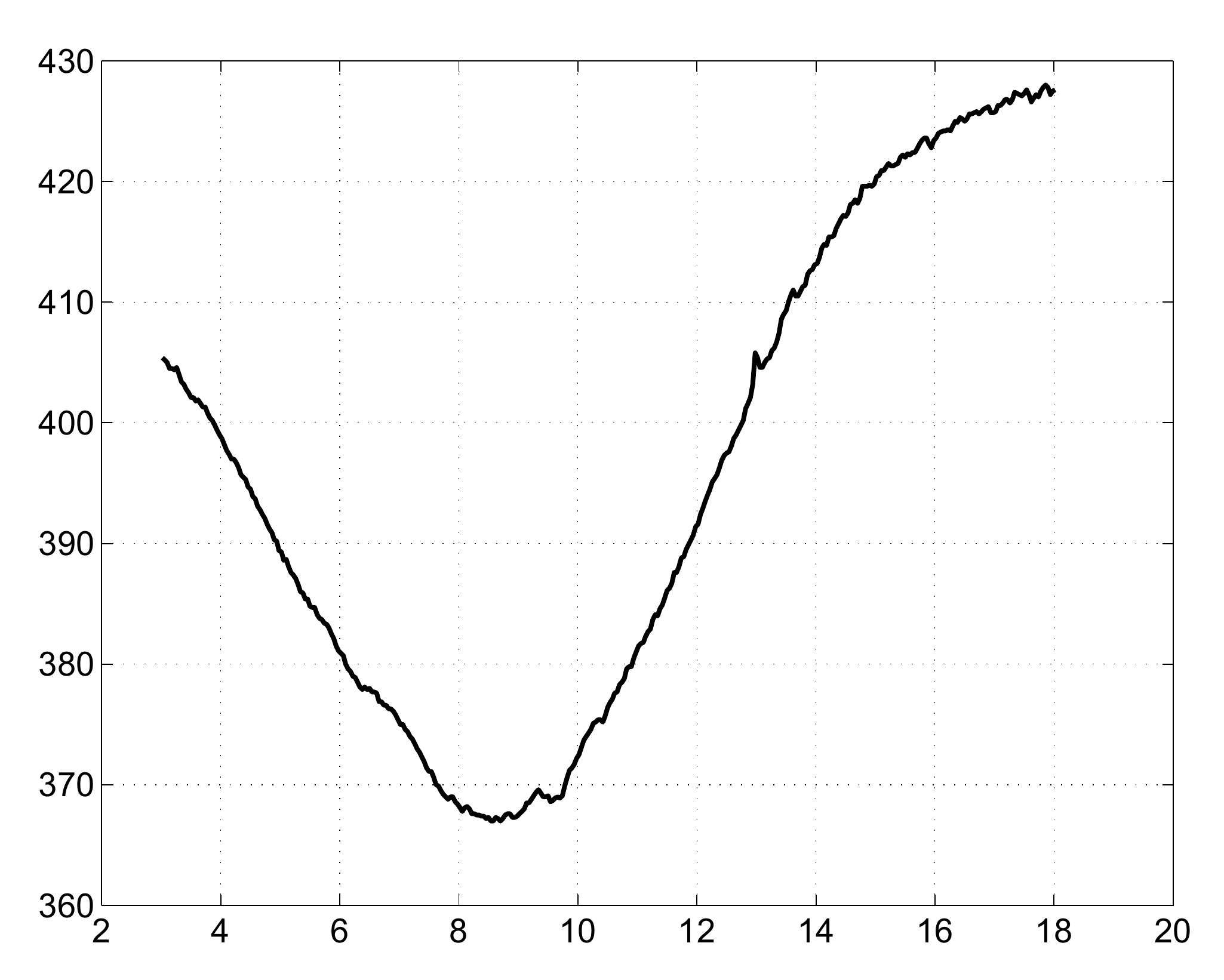}
\caption{Measurement input ($V_m$ in ft/s) versus time (s)}
\label{case3_vel}
\end{figure}

\begin{figure}[h]
\includegraphics[width=6in,height=2in]{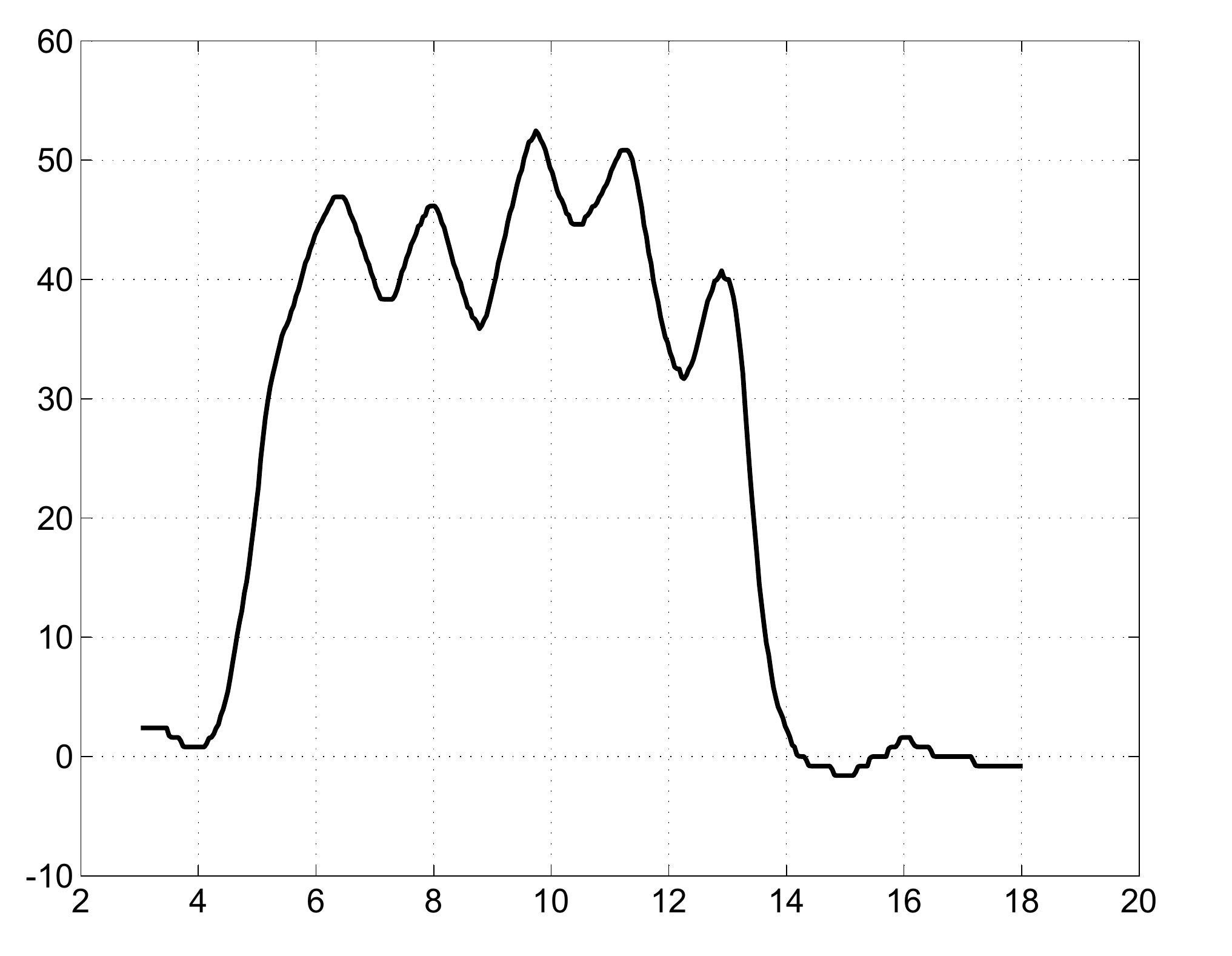}
\caption{Measurement input ($p_m$ in degrees) versus time (s)}
\label{case3_p}
\end{figure}

\begin{figure}[h]
\includegraphics[width=6in,height=2in]{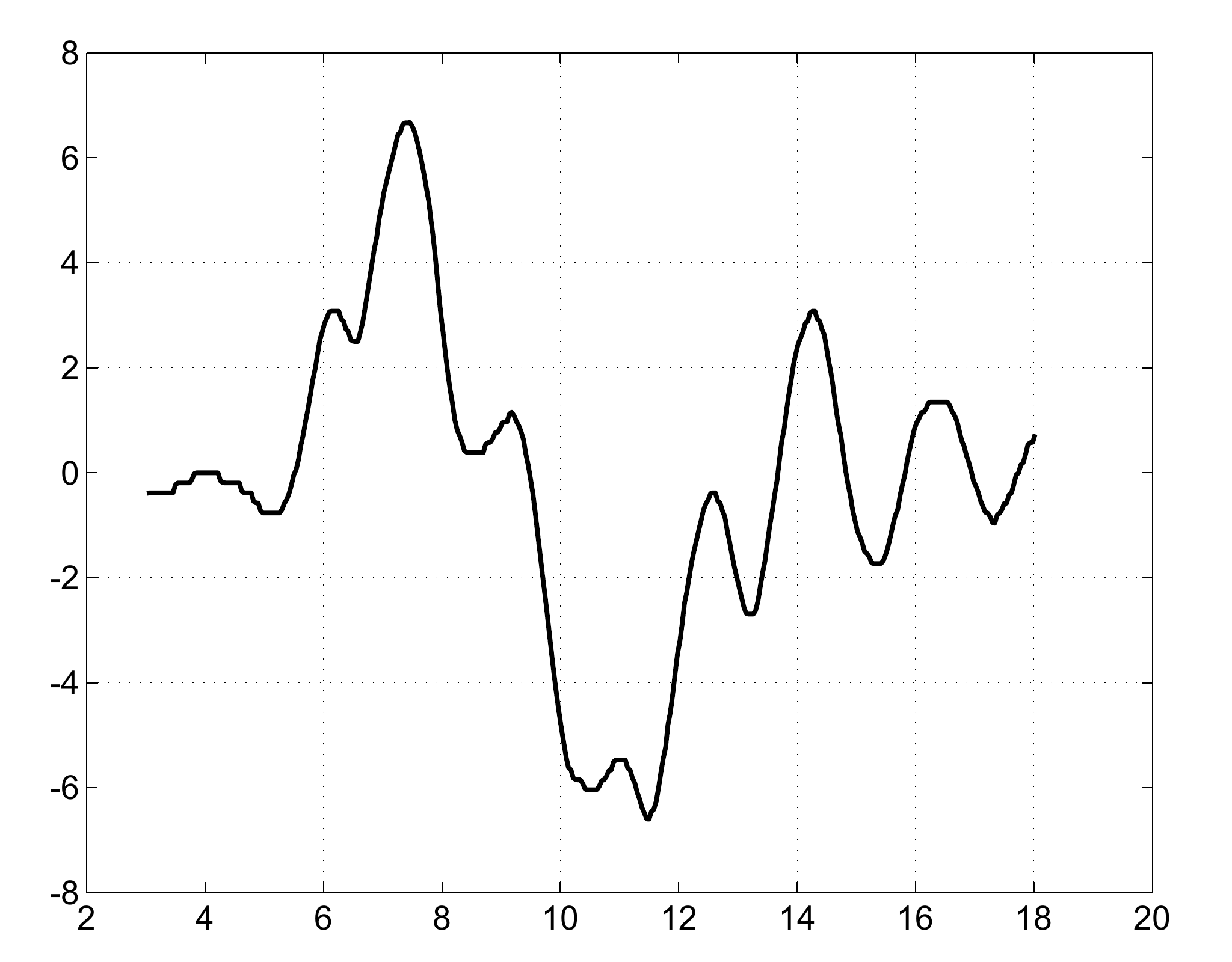}
\caption{Measurement input ($r_m$ in degrees) versus time (s)}
\label{case3_r}
\end{figure}

\begin{figure}[h]
\includegraphics[width=6in,height=5in]{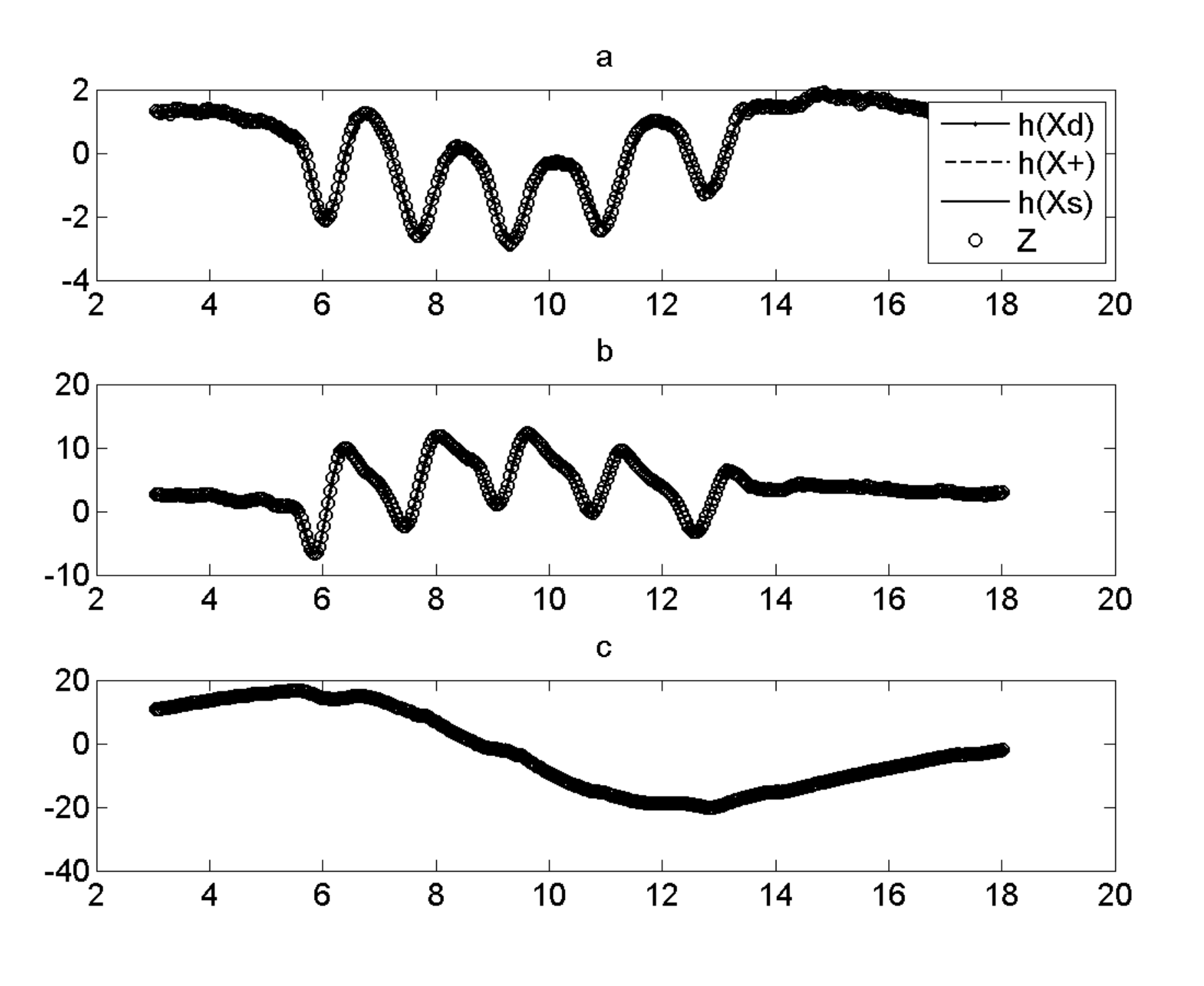}
\caption{Comparison of the predicted dynamics, posterior, smoothed}
\caption*{and the measurement in degrees(a. angle of attack b. pitch rate c. pitch angle) vs time}
\label{realQ3_s1}
\end{figure}

 The data set is obtained from NASA TP 1690 (Maine 1981) by employing a peculiar  manoeuvre  where elevator control input ($\delta_e$ in degrees) shown in Fig.\ref{input3} is imparted when the aircraft (T 37 B) is rolling through a full rotation about its x-axis during aileron roll. Similar to the earlier case, the coupling between the longitudinal and lateral motion are replaced by their measured values which includes roll angle ($\phi_m$), sideslip ($\beta_m$), velocity ($V_m$), roll rate ($p_m$), yaw rate ($r_m$) and the angle of attack ($\alpha_m$) as shown in Fig.\ref{case3_beta}, Fig.\ref{case3_vel}, Fig.\ref{case3_p}, Fig.\ref{case3_r}, Fig.\ref{case3_phi} and Fig.\ref{realQ3_s1} respectively. The state equations ($n=3$) for the angle of attack ($\alpha$), pitch rate (q) and the pitch angle ($\theta$) respectively are 
\begin{align*}
\dot{\alpha}=&-\frac{\bar qS}{mV_mcos(\beta_m)}(C_{L_\alpha}\alpha+C_{L_{\delta_e}} \delta_e+C_{L_0})+q+\frac{g}{V_mcos(\beta_m)}(cos(\phi_m)cos(\alpha_m)cos(\theta)+\\
&sin(\alpha_m)sin(\theta))-tan(\beta_m)(p_mcos(\alpha_m)+r_msin(\alpha_m))\\
\dot{q}=&\frac{\bar q S \bar c}{Iyy}(C_{m_\alpha}\alpha+C_{m_q}\frac{\bar c}{2V}q+C_{m_{\dot \alpha}}\frac{\bar c}{2V}\dot \alpha+C_{m_{\delta_e}}\delta_e+C_{m_0})+
\frac{Izz-Ixx}{Iyy}r_mp_m\\
\dot{\theta}=&qcos(\phi_m)-r_msin(\phi_m)+\theta_0
\end{align*}
The measurement equations ($m$=4) are given by
\begin{align*}
{\alpha_m}&=K_\alpha\alpha-K_\alpha x_\alpha \frac{q}{V} \\
{q_m}&=q\\
{\theta_m}&=\theta\\
{a_{n_m}}&=\frac{\bar q S}{mg}(C_{N_\alpha}\alpha+C_{N_{\delta_e}}\delta_e+C_{N_0})+\frac{x_{a_n}}{g}\dot{q} 
\end{align*}

The unknown parameters ($p=10$) is $(C_{L_{\alpha}},C_{L_{\delta_e}},C_{L_{0}},C_{m_{\alpha}},C_{m_{q}},C_{m_{\dot \alpha}},C_{m_{\delta_e}},C_{m_{0}},\theta_0,C_{N_0})^T$ with an approximation $C_{N_\alpha}=C_{L_\alpha}$ and $C_{N_{\delta_e}}=C_{L_{\delta_e}}$. The ones with suffix `$\delta_e$' are the control derivatives, the ones with suffix zero are the biases and all others are aerodynamic derivatives. The initial states are taken as initial measurement and the initial parameter values are taken as $(4,0.15,0.2,-0.5,-11.5,-5,-1.38,-0.06,-0.01,0.2)^T$.

\begin{table}[h]
\begin{center}
\caption*{Other constant values used for case-2}{}
\begin{tabular}{| c | c | c | c | c | c | }
\hline
 S=184 & m=196 & Ixx=6892.7  &  Iyy=3953.2 & Izz=10416.4  \\ \hline
g=32.2 & $\bar c$=5.58 & $K_\alpha x_\alpha$=-0.0279 & $x_{a_n}$=0.101 & $K_\alpha=1$\\ \hline
\end{tabular}
\end{center}
\end{table}

\par
\begin{figure}[h]
\includegraphics[width=6in,height=3in]{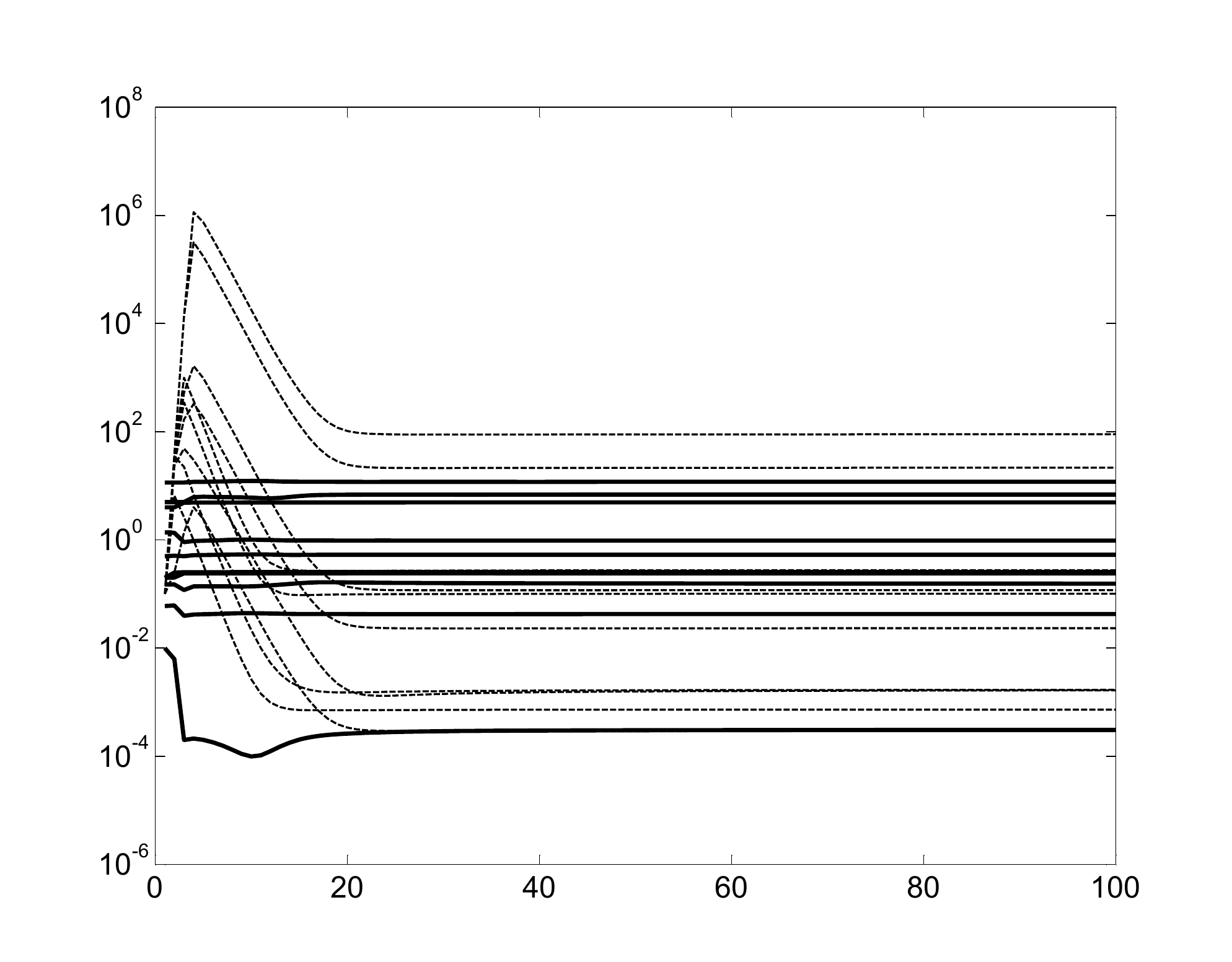}
\caption{Variation of initial parameters $\Theta_0$(continuous) and its $\mathbf{P_0}$(dashed) with iterations}
\label{realQ3_P0}
\end{figure}

\begin{figure}[h]
\includegraphics[width=6in,height=2.5in]{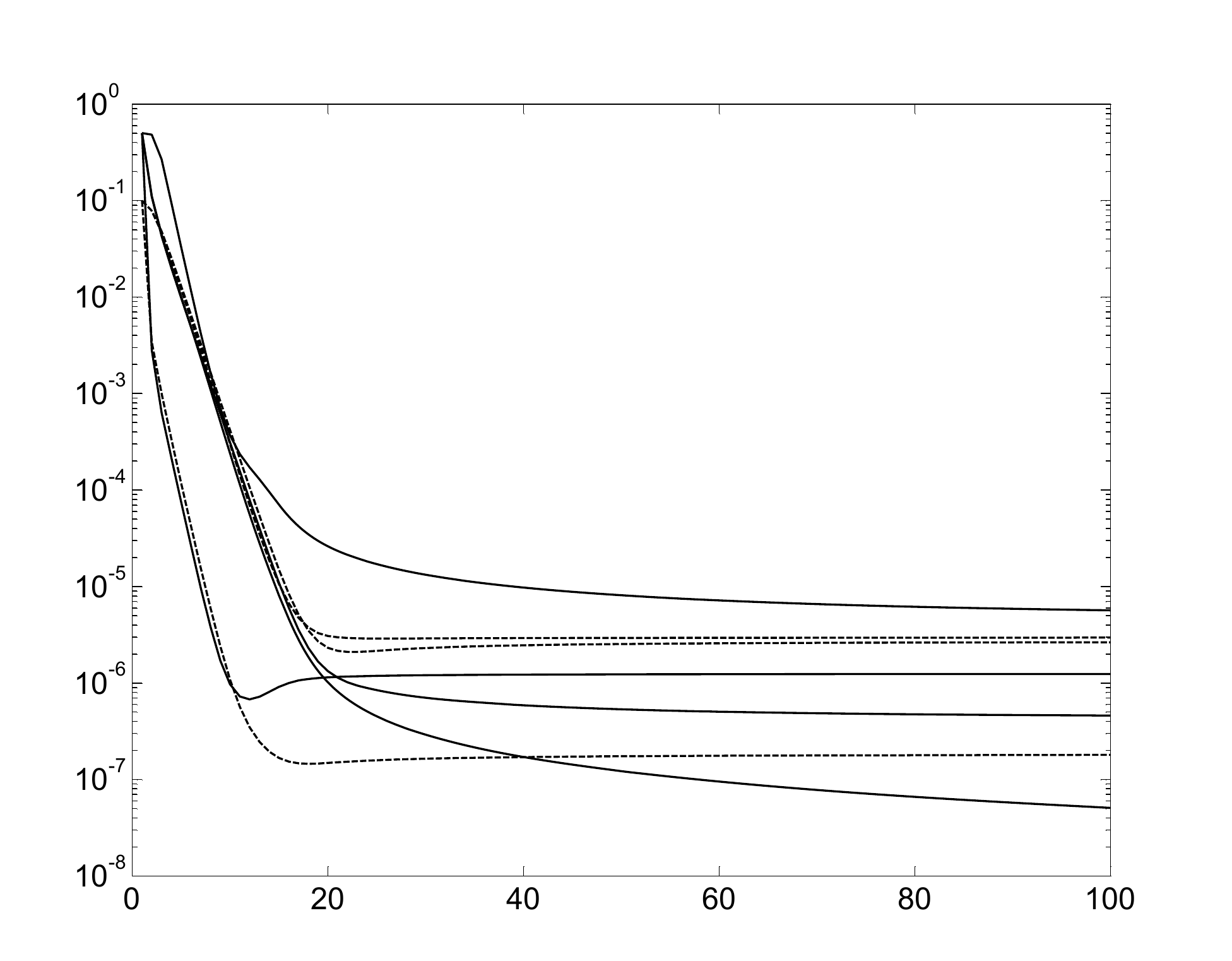}
\caption{Variation of \textbf{Q} (dashed) and \textbf{R} (continuous) with iterations}
\label{realQ3_R}
\end{figure}

\begin{figure}[h]
\includegraphics[width=6in,height=2in]{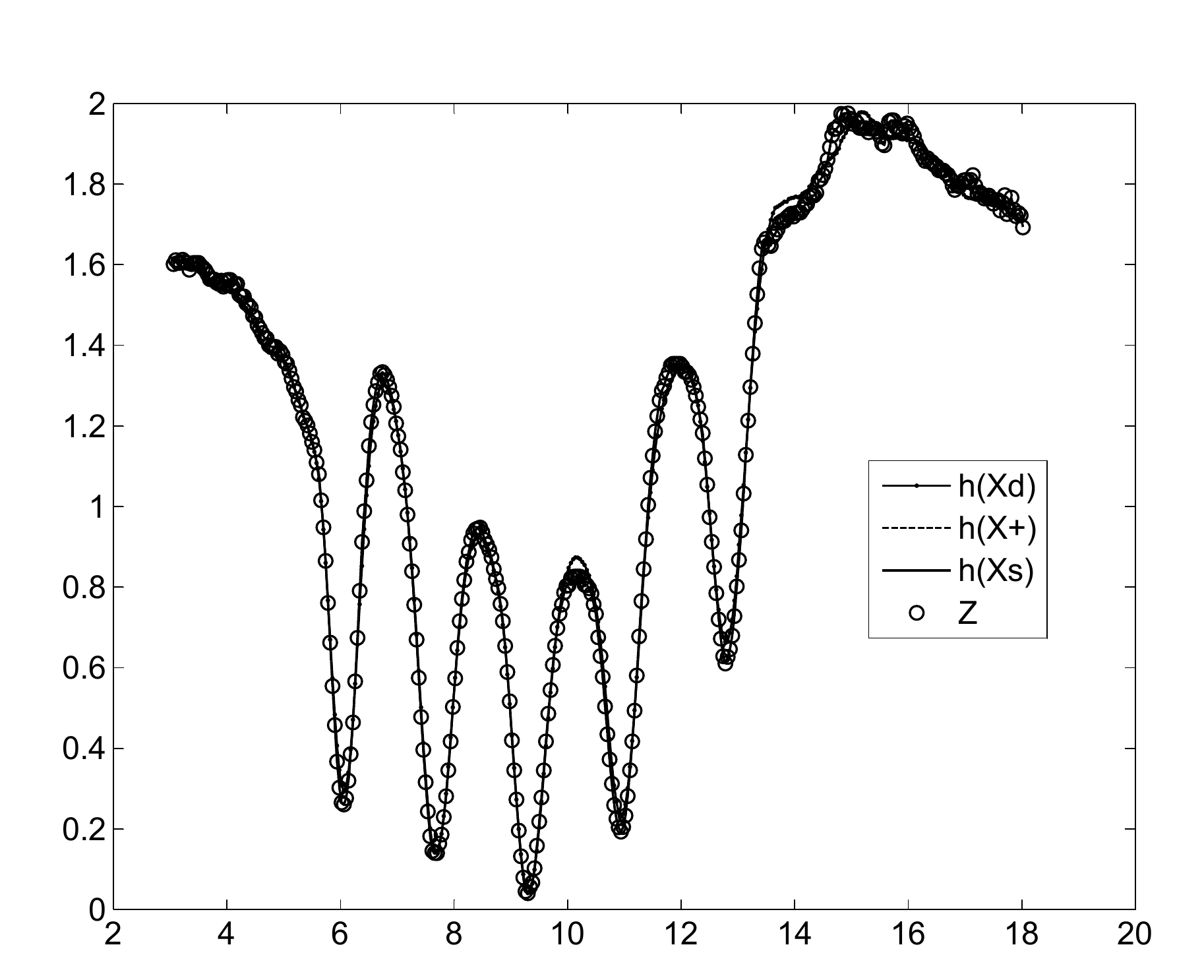}
\caption{Comparison of the predicted dynamics, posterior, smoothed}
\caption*{and the normal acceleration measurement in $ft/sec^2$ vs time}
\label{realQ3_h4}
\end{figure}

 Case-2 real data is run using the reference EKF (\textbf{Q} $>$ 0) with 100 iterations. The Figs.\ref{input3}-\ref{case3_r} are the inputs used in state equations. The Fig.\ref{realQ3_P0} shows the variation of parameter estimates and its initial covariance $\mathbf{P_0}$ with iterations and a similar Fig.\ref{realQ3_R} for \textbf{Q} and \textbf{R}. The values of \textbf{J1-J3} are close to the number of measurements ($m=4$) with \textbf{J6-J8} are close to the number of states ($n=3$) as shown in Fig.\ref{realQ3_J} and Table-\ref{tbcase3Q}. This means the measurement and state equations are well balanced. The \textbf{J5} is the negative log likelihood cost function. The later Figs.\ref{realQ3_s1}-\ref{realQ3_h4} compares (i) the state dynamics based on the estimated parameter after the filter pass through the data, (ii) the state after measurement update, (iii) the smoothed state and (iv) the measurement. Unlike in the simulated studies, the estimated measurement and process noise did not have constant statistical characteristics across time  Another experiment was carried out by generating a typical data set by using the estimated theta and injecting the estimated \textbf{Q} and \textbf{R} as additive white Gaussian noise. This is to determine the effect of  non white and non Gaussian noise distribution in the real data on the CRBs. After each iteration in the reference recipe the $\Theta$, \textbf{Q} and \textbf{R} were reset as from the real data. Similar experiment was also conducted by updating $\Theta$ as well. It turned out that there is not much of a difference in the final estimates and the CRBs. Two other filter runs were carried out using the MT and MS statistics for the estimation of \textbf{Q} and \textbf{R} with scaled up $\mathbf{P_0}$. The behaviour of the various cost function and in particular \textbf{J6} and \textbf{J7} in Table-\ref{tbcase3QMTMS} shows that the choice of the filter statistics for estimating \textbf{Q} and \textbf{R} in the proposed reference approach is the best possible when compared to other approaches.

\subsection{Remarks on Case - 2}
   
\par
\begin{figure}[h]
\includegraphics[width=6in,height=2.5in]{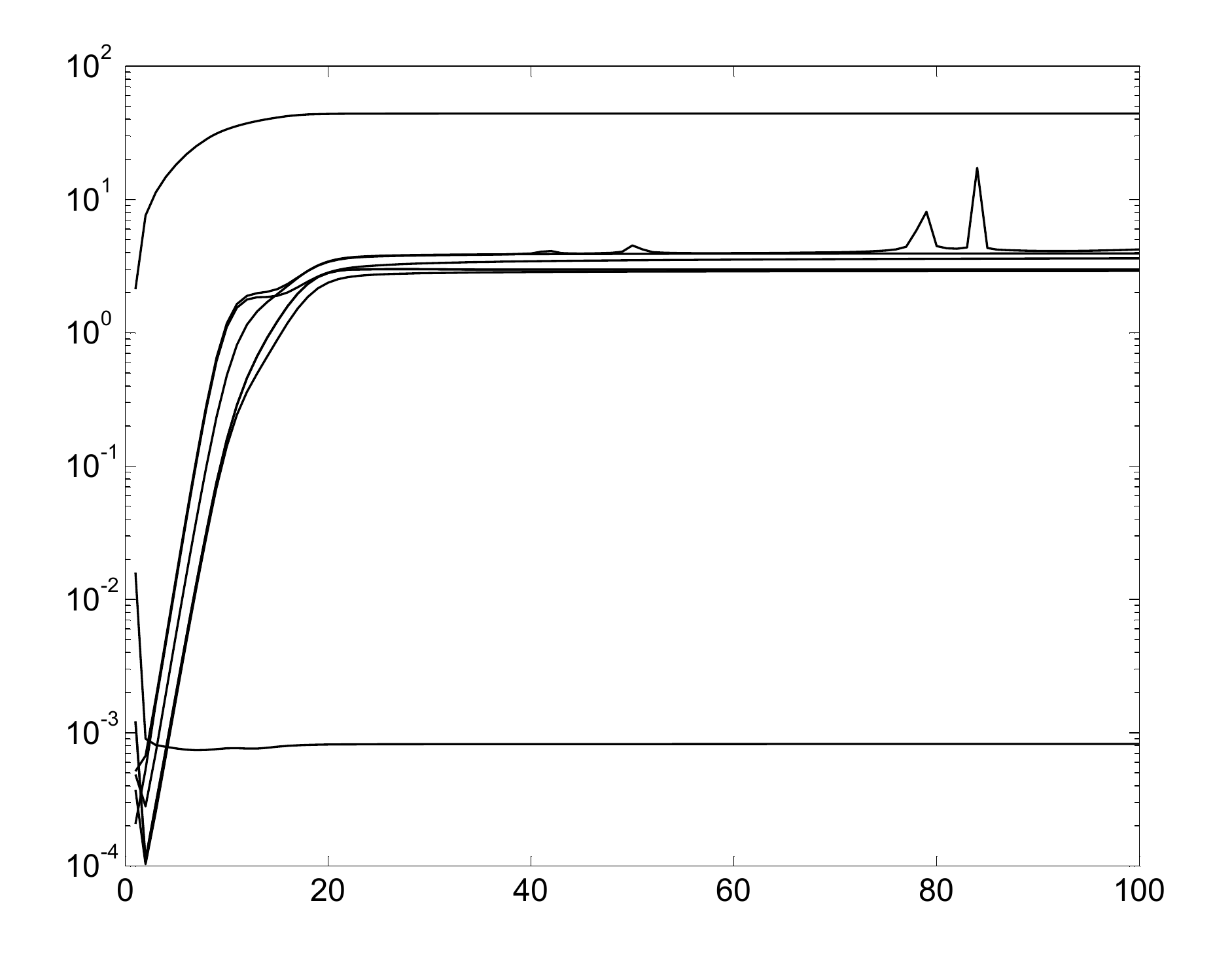}
\caption{Variation of different costs (\textbf{J1-J8}) with iterations}
\label{realQ3_J}
\end{figure}

\begin{figure}[h]
\includegraphics[width=6in,height=3in]{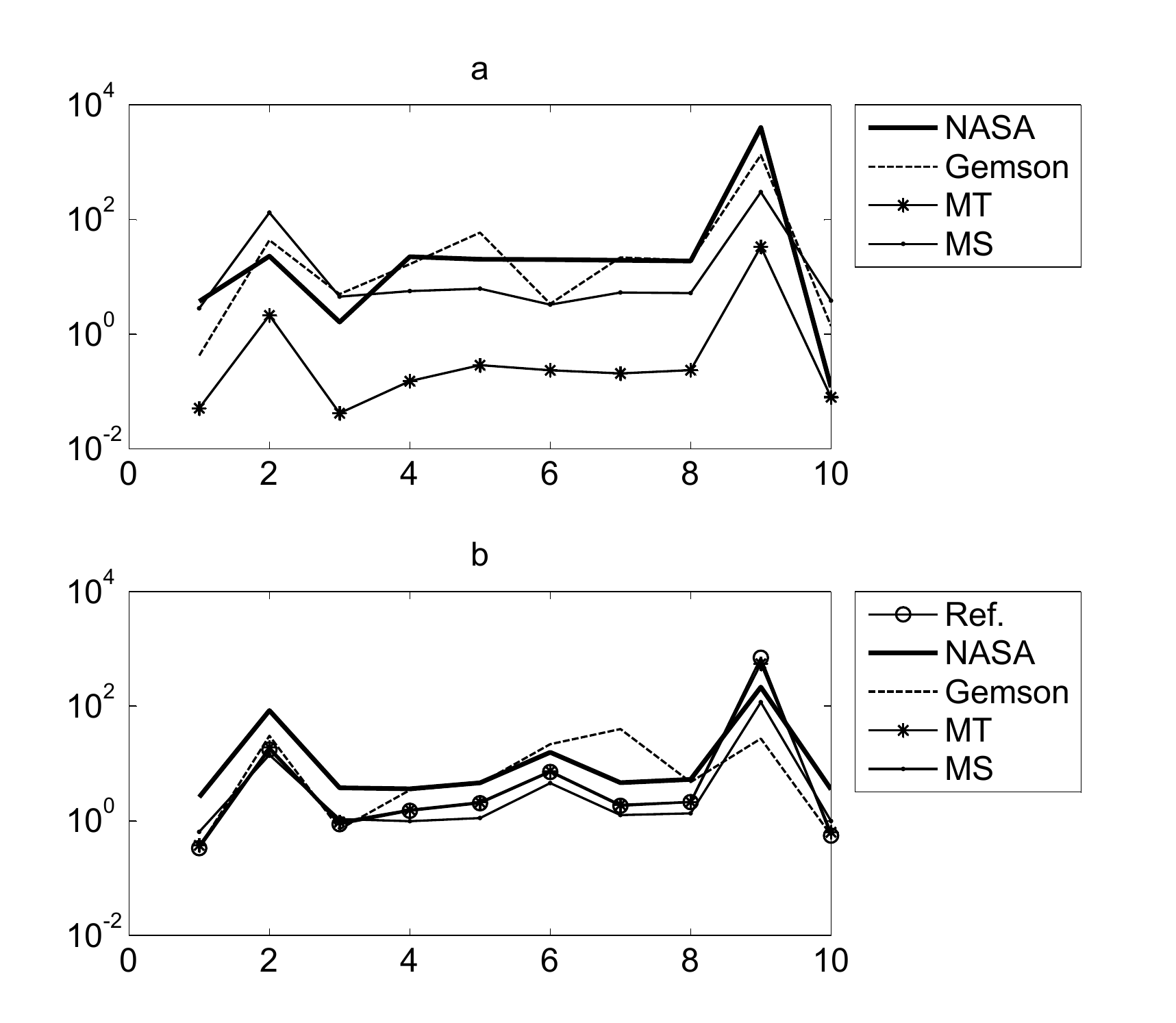}
\caption{Comparison of (a) Absolute percentage error with respect to the reference of the parameter estimates and (b) $\%$CRBs  by different methods}
\label{comp3}
\end{figure}

 The NASA results have been generated assuming \textbf{Q} = 0 and are comparable with reference procedure for the parameter estimates and their CRBs. Further the MT and MS methods give quite different estimates for the \textbf{Q} and \textbf{R} values than from the reference case. We believe that the reference procedure provides the best possible parameter estimates and their uncertainties. From the plot of \% error in the parameter estimate with respect to the reference value and the \%CRB in Fig \ref{comp3}, it can be seen that the parameters $C_{L_{\delta_e}}$ and $\theta_0$ are relatively weak when compared to other parameters. The CRBs as estimated by different methods generally appear to vary widely. However what is interesting is that even the estimate of the strong parameter such as $C_{m_{q}}$ varies widely among the methods. Such a behaviour of the filter across the parameter estimates shows how important is the tuning of the filter statistics namely $\mathbf{P_0}$, \textbf{Q} and \textbf{R} in parameter estimation and their uncertainties. It was also observed that for this particular case, the cost \textbf{J2} using filtered residue went negative at some iteration as seen in Fig.\ref{realQ3_J} with a spike whose absolute value was plotted. However the cost \textbf{J3} using smoothed residue that was used for tuning the filter did not show any such peculiarity. The rounded off 100$\times$C matrix of the parameter estimates for this case is 
\begin{center}
\begin{tabular}{|c|c|c|c|c|c|c|c|c|c|c|c|c|c|}
\hline

$\Theta$ & $C_{L_{\alpha}}$ & $C_{L_{\delta_e}}$ & $C_{L_{0}}$ & $C_{m_{\alpha}}$ & $C_{m_{q}}$ & $C_{m_{\dot \alpha}}$ & $C_{m_{\delta_e}}$ & $C_{m_{0}}$ & $\theta_0$ & $C_{N_0}$ \\ \hline 
$C_{L_{\alpha}}$ &  100	&	67	&	41	&	-19	&	1	&	1	&	-8	&	-8	&	0	&	62	\\ \hline
$C_{L_{\delta_e}}$ & 67	&	100	&	64	&	-11	&	2	&	1	&	-12	&	-13	&	0	&	98	\\ \hline
$C_{L_{0}}$ & 41	&	64	&	100	&	-4	&	-2	&	5	&	-5	&	-7	&	0	&	65	\\ \hline
$C_{m_{\alpha}}$ & -19	&	-11	&	-4	&	100	&	25	&	70	&	91	&	88	&	0	&	-10	\\ \hline
$C_{m_{q}}$ & 1	&	2	&	-2	&	25	&	100	&	-27	&	21	&	9	&	1	&	1	\\ \hline
$C_{m_{\dot \alpha}}$ &  1	&	1	&	5	&	70	&	-27	&	100	&	80	&	84	&	-1	&	2	\\ \hline
$C_{m_{\delta_e}}$ & -8	&	-12	&	-5	&	91	&	21	&	80	&	100	&	99	&	0	&	-12	\\ \hline
$C_{m_{0}}$ & -8	&	-13	&	-7	&	88	&	9	&	84	&	99	&	100	&	0	&	-13	\\ \hline
$\theta_0$ & 0	&	0	&	0	&	0	&	1	&	-1	&	0	&	0	&	100	&	0	\\ \hline
$C_{N_0}$ & 62	&	98	&	65	&	-10	&	1	&	2	&	-12	&	-13	&	0	&	100	\\ \hline
\end{tabular}
\end{center}

\begin{table}[h]
\caption{Real flight test data case-2 results ($\Theta,\sigma_{\Theta}$).}{}
\label{tbcase3Q}
\begin{center}
\begin{tabular}{|c| c| c| c| c| c|c|c|c|c|c| }
\hline
$\Theta$ & 
Reference &   
\makecell{NASA} &   
\makecell{Gemson} &  
MT &
MS
\\ \hline

\makecell{ $C_{L_{\alpha}}$ \\ \\ $C_{L_{\delta_e}}$ \\ \\ $C_{L_{0}}$ \\ \\ $C_{m_{\alpha}}$ \\ \\ $C_{m_{q}}$ \\ \\ $C_{m_{\dot \alpha}}$ \\ \\ $C_{m_{\delta_e}}$ \\ \\ $C_{m_{0}}$ \\ \\ $\theta_0$ \\ \\ $C_{N_0}$ } &

\makecell{ 4.9235  \\ (0.0164) \\ 0.1554  \\  (0.0271) \\ 0.2409  \\(0.0021)  \\ -0.5293\\  (0.0079)  \\ -11.8596 \\  (0.2402) \\ -6.8959  \\ (0.4891)  \\ -0.9731 \\(0.0177) \\ -0.0425  \\(0.0009) \\  0.0003   \\  (0.0021)   \\0.2538 \\ (0.0014)} &

\makecell{ 5.1068 \\ (0.1322)  \\  0.1909  \\  (0.1602) \\ 0.2448  \\ (0.009215) \\ -0.6474 \\ (0.02339)  \\-14.26  \\  (0.6528) \\ -8.27 \\ (1.296) \\ -1.1614 \\ (0.05371)  \\ -0.0505 \\ (0.002655) \\ -0.01177 \\  (0.02528)  \\ 0.2541 \\(0.008935) } &

\makecell{ 4.9028 \\ (0.0168)  \\  0.0879  \\  (0.0267) \\  0.2529  \\(0.0018) \\ -0.6174 \\ (0.0211)  \\ -18.8339 \\(0.8379) \\ -7.1290 \\(1.544) \\ -1.1841 \\ (0.471)  \\ -0.0507 \\(0.0024) \\ -0.0037 \\ (0.001)  \\ 0.2503 \\ (0.0014) } &

\makecell{   4.9260 \\ (0.0184) \\  0.1587 \\ (0.0302) \\  0.2408 \\  (0.0023) \\ -0.5285 \\ (0.0082)  \\-11.8255 \\ (0.2483)  \\ -6.8798  \\ (0.5062) \\ -0.9711 \\ (0.0184) \\ -0.0424 \\ (0.0009) \\  0.0002  \\ (0.0011) \\ 0.2540 \\ (0.0016)} &

\makecell{  5.0620 \\(0.0323) \\  0.3594 \\  (0.0508)  \\  0.2517 \\  (0.0027) \\  -0.5590 \\  (0.0055) \\ -12.5965 \\  (0.1400) \\  -6.6713 \\   (0.3021) \\  -1.0247 \\  (0.0129) \\ -0.0447 \\ (0.0006) \\ -0.0006 \\   (0.0007)  \\ 0.2635 \\ (0.0026) }

\\ \hline

\end{tabular}
\end{center}

\vspace{0.5cm}

\caption{Real flight test data case-2 results* (\textbf{R,Q,J}).}{}
\label{tbcase3QMTMS}
\begin{center}
\begin{tabular}{|c| c| c|| c| c| c|| c|c|c| }
\hline

\makecell{\textbf{R} (Ref)\\ $\times10^{-6}$ }&  
\makecell{\textbf{Q} (Ref)\\ $\times10^{-6}$}&  
\makecell{\textbf{J1-J8} \\(Ref) }&

\makecell{\textbf{R} (MT)\\ $\times10^{-6}$ }&  
\makecell{\textbf{Q} (MT)\\ $\times10^{-6}$}&  
\makecell{\textbf{J1-J8} \\(MT) }&

\makecell{\textbf{R} (MS)\\ $\times10^{-6}$ }&  
\makecell{\textbf{Q} (MS)\\ $\times10^{-6}$}&  
\makecell{\textbf{J1-J8} \\(MS) }

\\ \hline

\makecell{ 1.241  \\  0.051 \\   0.460 \\   5.668} &
\makecell{ 0.180  \\  2.954  \\  2.646} &
\makecell{ 3.9336   \\ 4.2225  \\  3.6162  \\  0.0008  \\ -44.1347  \\  2.9752 \\   2.9760 \\ 2.9070} &

\makecell{ 1.6135   \\ 0.2395  \\  2.3155  \\  2.9290} &
\makecell{0.2025 \\   3.1532  \\  0.6666} &
\makecell{ 3.7662  \\  4.5191  \\  3.8384  \\  0.0008 \\ -43.7340  \\  4.2266  \\  4.2284 \\ 2.9489} &

\makecell{ 3.1599 \\  37.2424 \\   9.3413 \\ 841.5496} &
\makecell{  0.00005  \\  0.0003  \\  0.2386} &
\makecell{  3.1621   \\ 3.1507  \\  2.5900 \\   0.0007 \\ -38.0517  \\  8.4768  \\  8.4655 \\ 3.0215}

\\ \hline

\end{tabular}
*Cost functions are not close to their expected values in MT and MS methods.
\end{center}
\end{table}

\clearpage
\section{Analysis of Real Flight Test Case - 3}
\par
\begin{figure}[h]
\includegraphics[width=6in,height=2in]{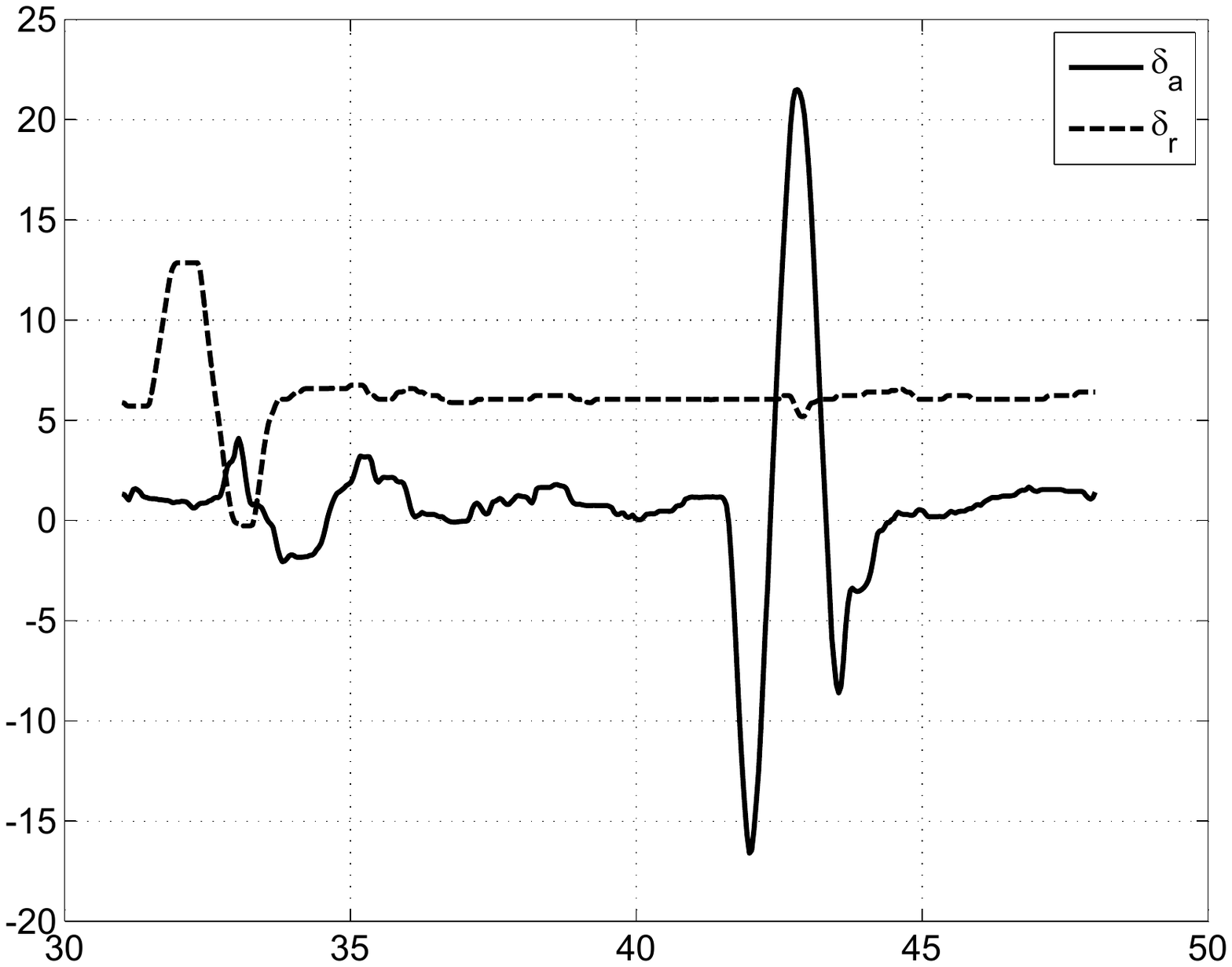}
\caption{Control input ($\delta_a,\delta_r$ in degrees) versus time (s)}
\label{input5}
\end{figure}

\begin{figure}[h]
\includegraphics[width=6in,height=2in]{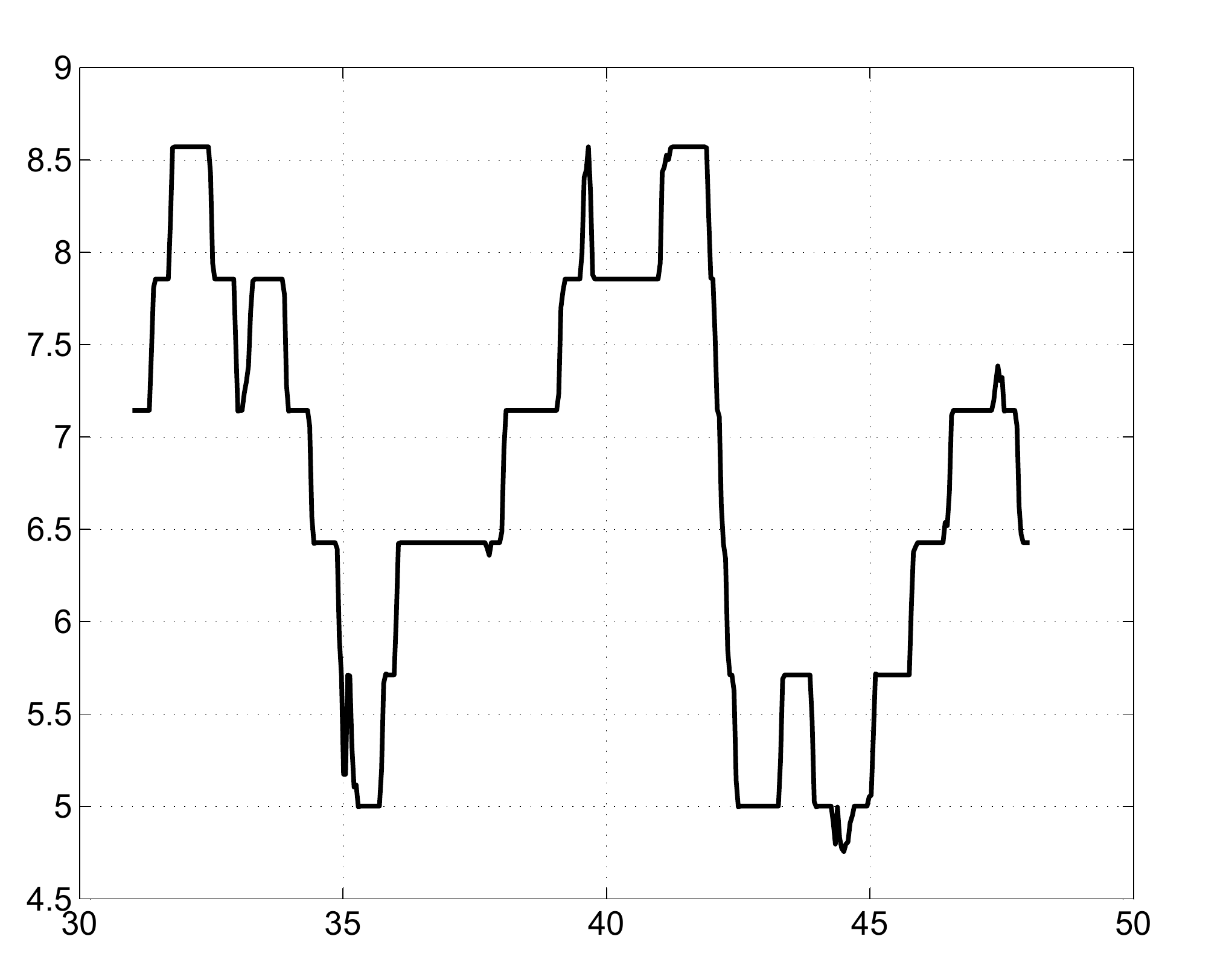}
\caption{Measurement input ($\theta_m$ in degrees) versus time (s)}
\label{case5_theta}
\end{figure}

\begin{figure}[h]
\includegraphics[width=6in,height=2in]{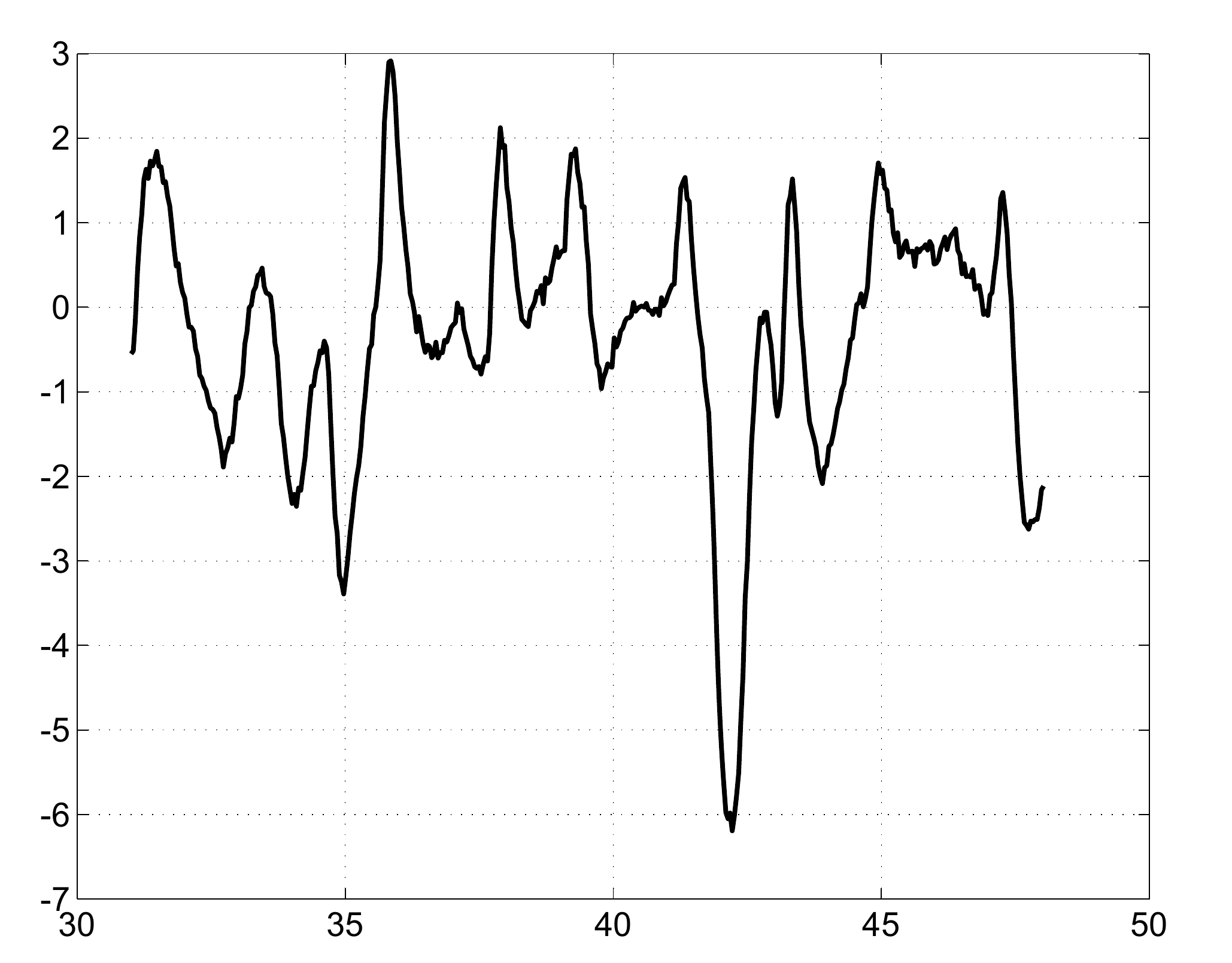}
\caption{Measurement input ($q_m$ in degrees) versus  time (s)}
\label{case5_q}
\end{figure}

\begin{figure}[h]
\includegraphics[width=6in,height=2in]{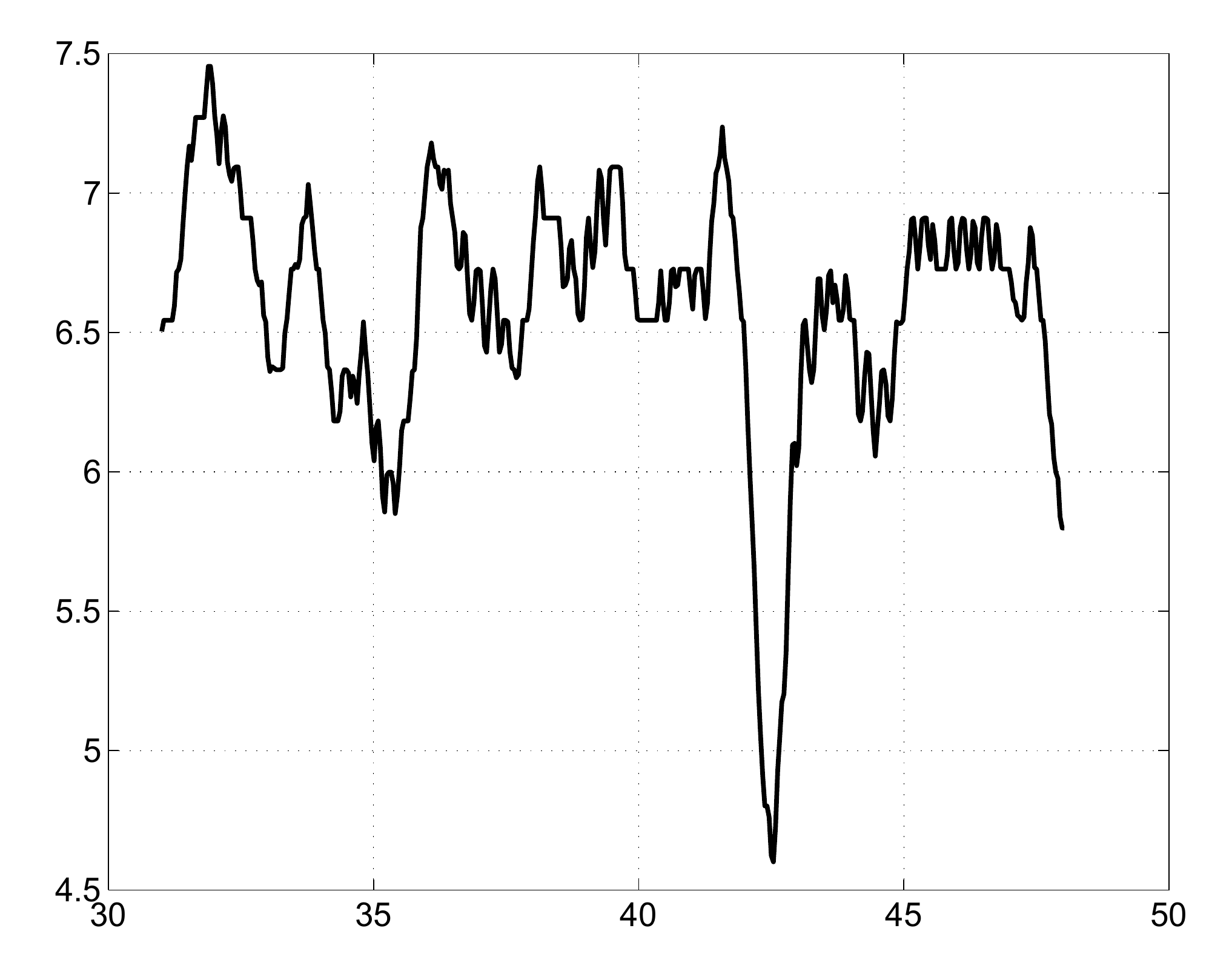}
\caption{Measurement input ($\alpha_m$ in degrees) versus  time (s)}
\label{case5_alpha}
\end{figure}

 The data set is obtained from NASA TP 1690 which describes the lateral motion of a oblique wing aircraft with zero wing skew excited by the control input ($\delta_a$ and $\delta_r$ in degrees) as shown in Fig.\ref{input5}. Similar to the earlier case, the coupling between the longitudinal and lateral motion are replaced by their measured values which includes pitch angle ($\theta_m$ in rad), pitch rate ($q_m$ in rad/s) and the angle of attack ($\alpha_m$ in rad) as shown in Fig.\ref{case5_theta}, Fig.\ref{case5_q} and Fig.\ref{case5_alpha} respectively. The state equations ($n$=4) for the angle of sideslip ($\beta$), roll rate (p), roll angle ($\phi$) and yaw rate (r) are
\begin{align*}
\dot{\beta}=&\frac{\bar qS}{mV}(C_{Y_\beta}\beta+C_{Y_p}\frac{b}{2V}p+C_{Y_r}\frac{b}{2V}r+C_{Y_{\delta_a}} \delta_a+C_{Y_{\delta_r}} \delta_r+\beta_0)+\\&\frac{g}{V}sin(\phi)cos(\theta_m)+psin(\alpha_m)-rcos(\alpha_m) \\
\dot{p}-\dot{r}\frac{Izx}{Ixx}=&\frac{\bar q S b}{Ixx}(C_{L_\beta}\beta+C_{L_p}\frac{\bar c}{2V}p+C_{L_r}\frac{\bar c}{2V}r+C_{L_{\delta_a}} \delta_a+C_{L_{\delta_r}} \delta_r+C_{L_0})+\\&\frac{Iyy-Izz}{Ixx}rq_m+\frac{Izx}{Ixx}pq_m\\
\dot{\phi}=&p+q\text{ }tan(\theta_m)sin(\phi)+r\text{ }tan(\theta_m)cos(\phi)+\phi_0\\
\dot{r}-\dot{p}\frac{Izx}{Izz}=&\frac{\bar q S b}{Izz}(C_{N_\beta}\beta+C_{N_p}\frac{b}{2V}p+C_{N_r}\frac{b}{2V}r+C_{N_{\delta_a}} \delta_a+C_{N_{\delta_r}} \delta_r+C_{N_0})+\\&\frac{Ixx-Iyy}{Izz}pq_m-\frac{Izx}{Izz}rq_m
\end{align*}
The measurement equations ($m$=5) are given by
\begin{align*}
{\beta_m}&=\beta-K_\beta z_\beta \frac{p}{V}+K_\beta x_\beta \frac{r}{V} \\
{p_m}&=p\\
{\phi_m}&=\phi\\
r_m&=r\\
{a_{y_m}}&=\frac{\bar q S}{mg}(C_{Y_\beta}\beta+C_{Y_p}\frac{b}{2V}p+C_{Y_r}\frac{b}{2V}r+C_{Y_{\delta_a}}\delta_a+C_{Y_{\delta_r}}\delta_r+C_{Y_0})-\frac{z_{a_y}}{g}\dot{p}+\frac{x_{a_y}}{g}\dot{r}
\end{align*}

The unknown parameter set ($p=20$) is $\Theta=(C_{Y_{\beta}},C_{Y_{\delta_r}},\beta_0,C_{L_{\beta}},C_{L_{p}},C_{L_r},C_{L_{\delta_a}},C_{L_{\delta_r}},C_{L_{0}}, \phi_0,C_{N_{\beta}},\\C_{N_{p}},C_{N_r},C_{N_{\delta_a}},C_{N_{\delta_r}},C_{N_0},C_{Y_0},C_{Y_p},C_{Y_r},C_{Y_{\delta_a}})^T$. The ones with suffix `$\delta_a$' and `$\delta_r$' are the control derivatives, the ones with suffix zero are the biases and all others are aerodynamic derivatives. The initial states are taken as initial measurement  and the initial parameter values are taken as $(-0.5,0.1,-0.01,0.01,-0.35,0.01,0.06,0.01,-0.002,0.002,0.07,-0.055,-0.05,0.003,-0.04,0.0068,\\-0.025,0.5,-1,0.005)^T$. 

\begin{table}[h]
\begin{center}
\caption*{Other constant values used for case-3}{}
\begin{tabular}{| c | c | c | c | c | c | c | }
\hline
$\bar q=865.3$ & S=9.3 & m=387.7 & Ixx=314  &  Iyy=488 & Izz=698 & Izx=69   \\ \hline
V=39.41 & g=9.81 & b=6.81 & $K_\beta z_\beta$=0.305 & $K_\beta x_\beta$=2.73 & $z_{a_y}$=-0.098 & $x_{a_y}$=0.651 \\ \hline
\end{tabular}
\end{center}
\end{table}

\par
\begin{figure}[h]
\includegraphics[width=6in,height=3in]{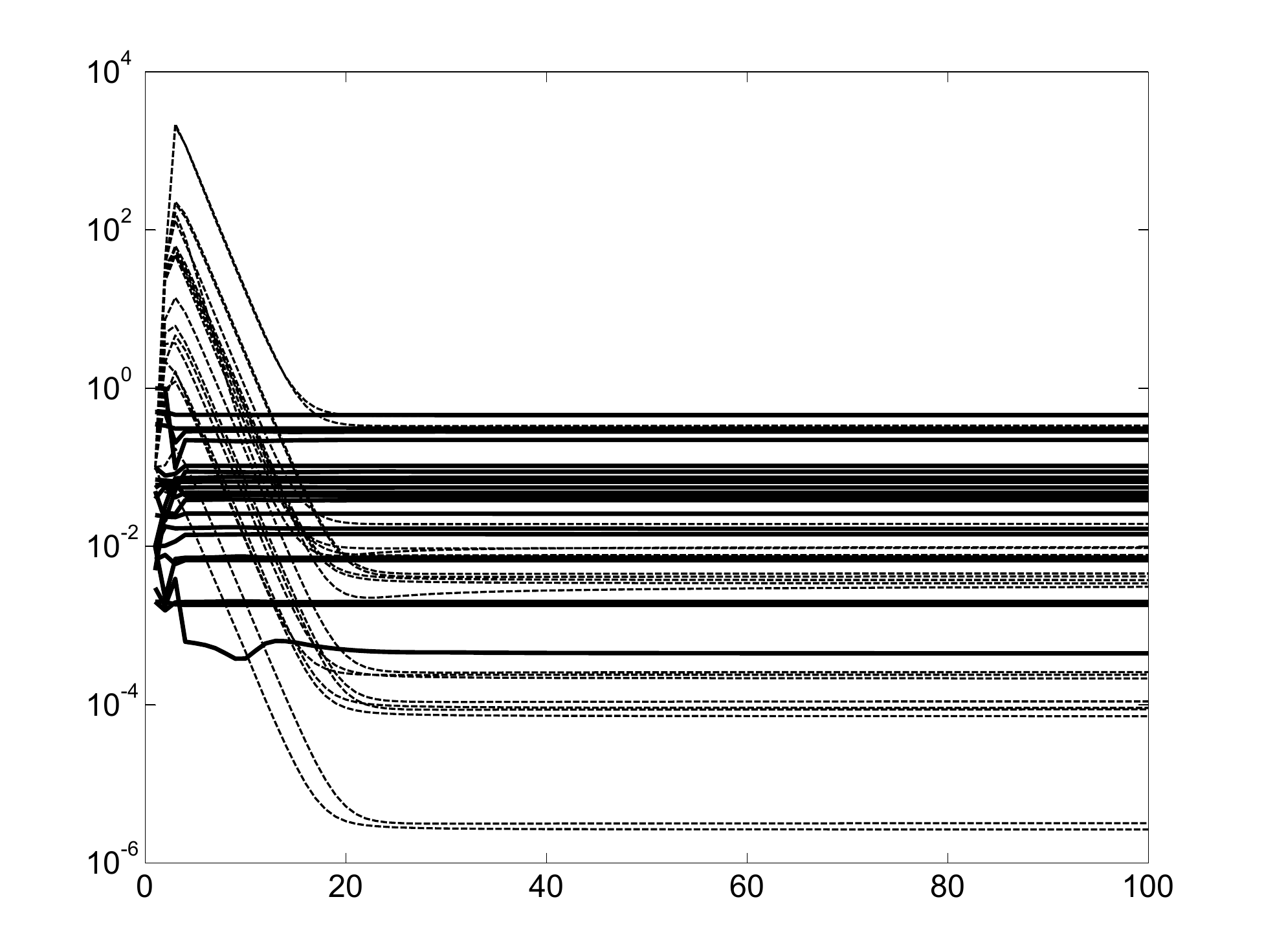}
\caption{Variation of initial parameters $\Theta_0$(continuous) and its $\mathbf{P_0}$(dashed) with iterations}
\label{realQ5_P0}
\end{figure}

\begin{figure}[h]
\includegraphics[width=6in,height=2.5in]{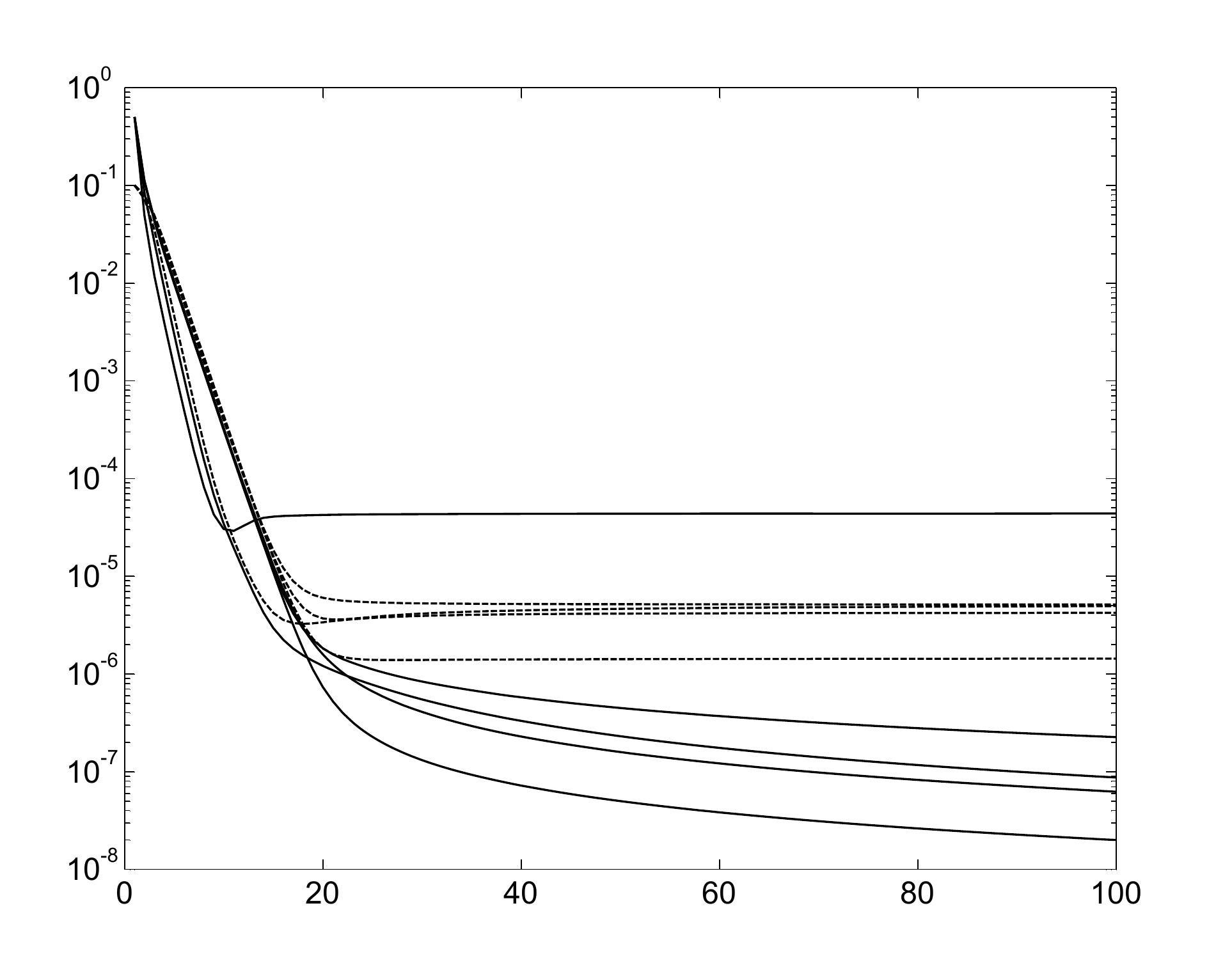}
\caption{Variation of \textbf{Q} (dashed) and \textbf{R}(continuous) with iterations}
\label{realQ5_R}
\end{figure}

\begin{figure}[h]
\includegraphics[width=6in,height=2.5in]{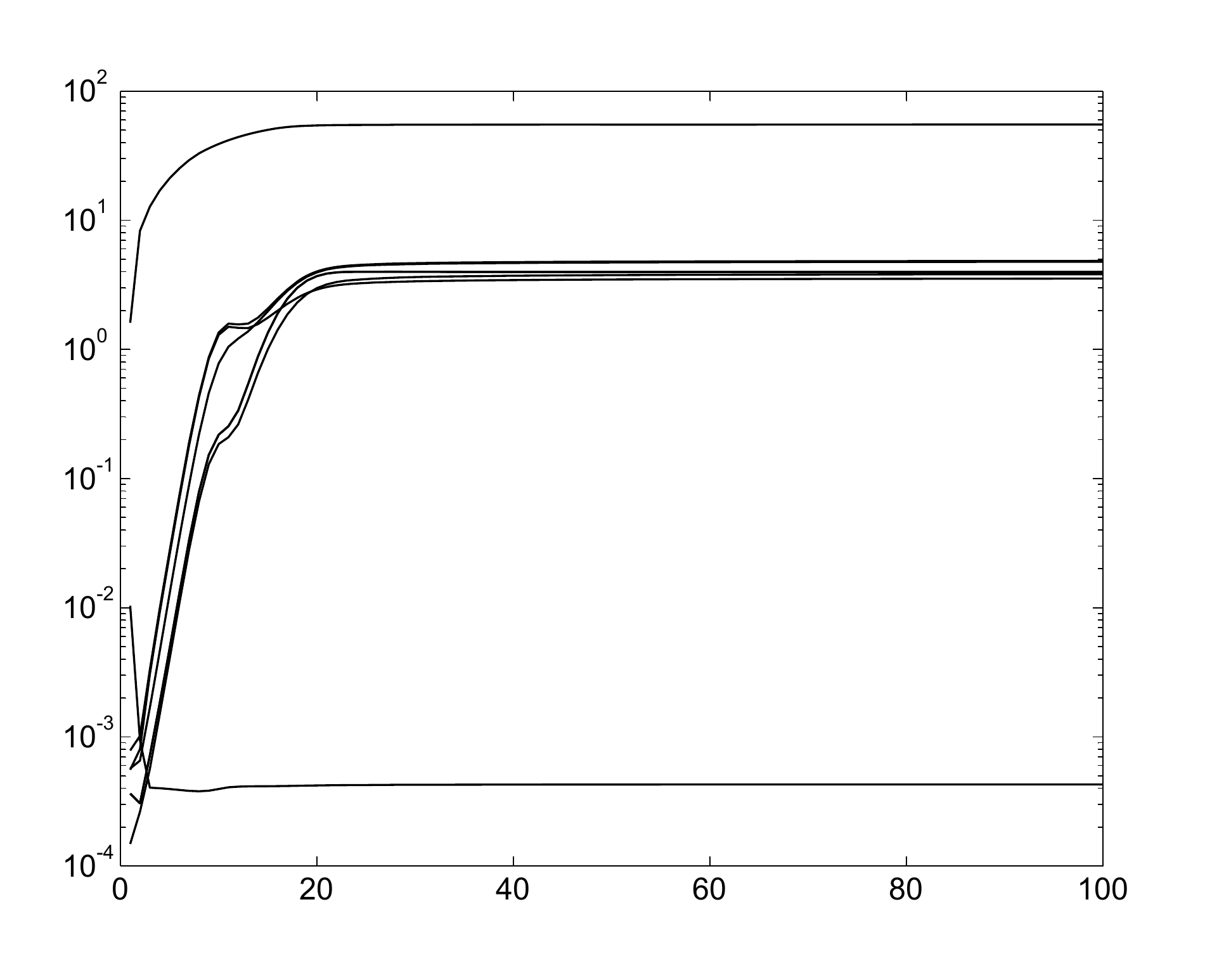}
\caption{Variation of different costs (\textbf{J1-J8}) with iterations}
\label{realQ5_J}
\end{figure}

\begin{figure}[h]
\includegraphics[width=6in,height=2in]{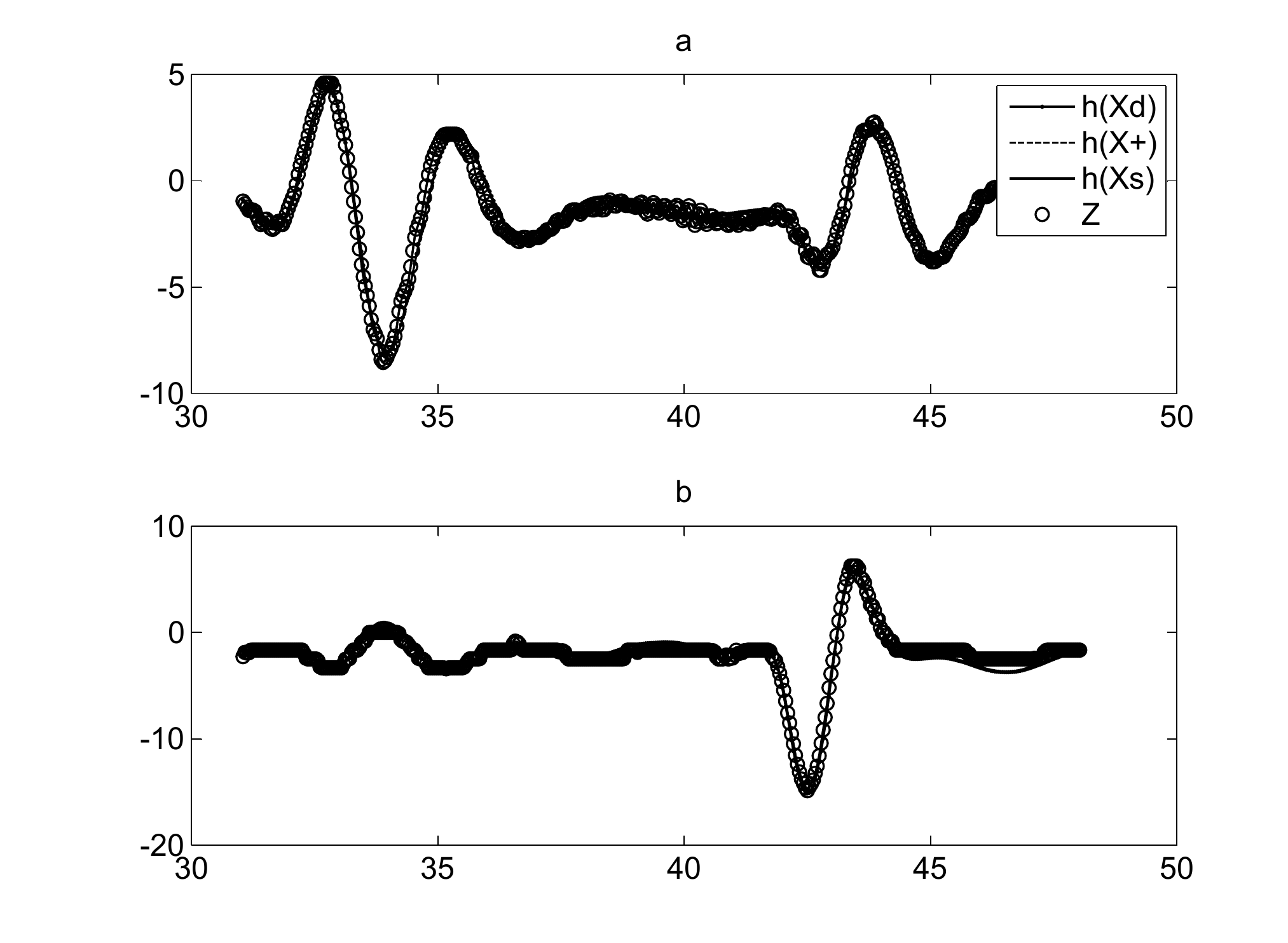}
\caption{Comparison of the predicted dynamics, posterior, smoothed}
\caption*{and the measurement in degrees (a. Sideslip b. Roll angle) vs time }
\label{realQ5_s1}
\end{figure}

\begin{figure}[h]
\includegraphics[width=6in,height=2.5in]{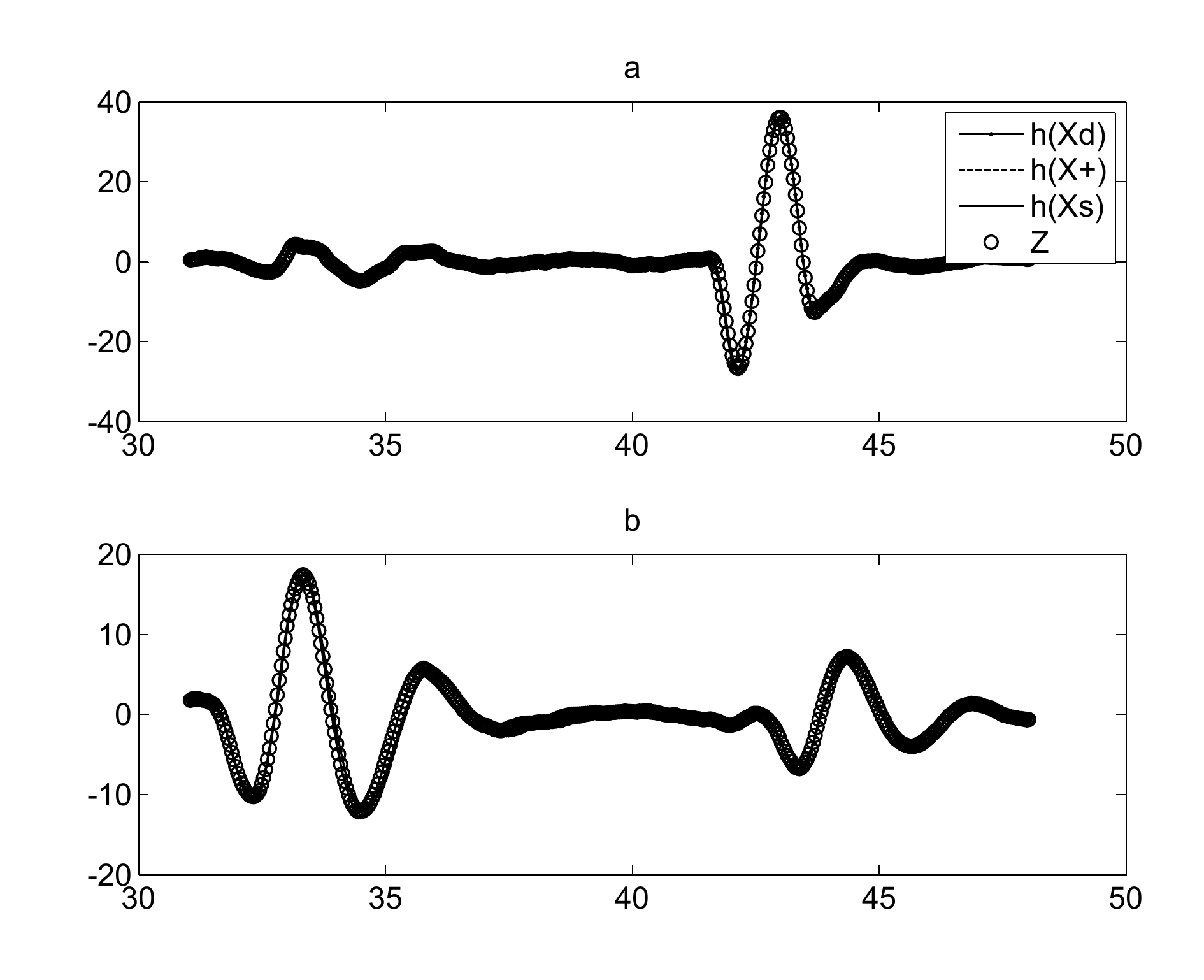}
\caption{Comparison of the predicted dynamics, posterior, smoothed}
\caption*{and the measurement in degrees/sec (a. Roll rate b. Yaw rate) vs time}
\label{realQ5_s2}
\end{figure}

\begin{figure}[h]
\includegraphics[width=6in,height=1.5in]{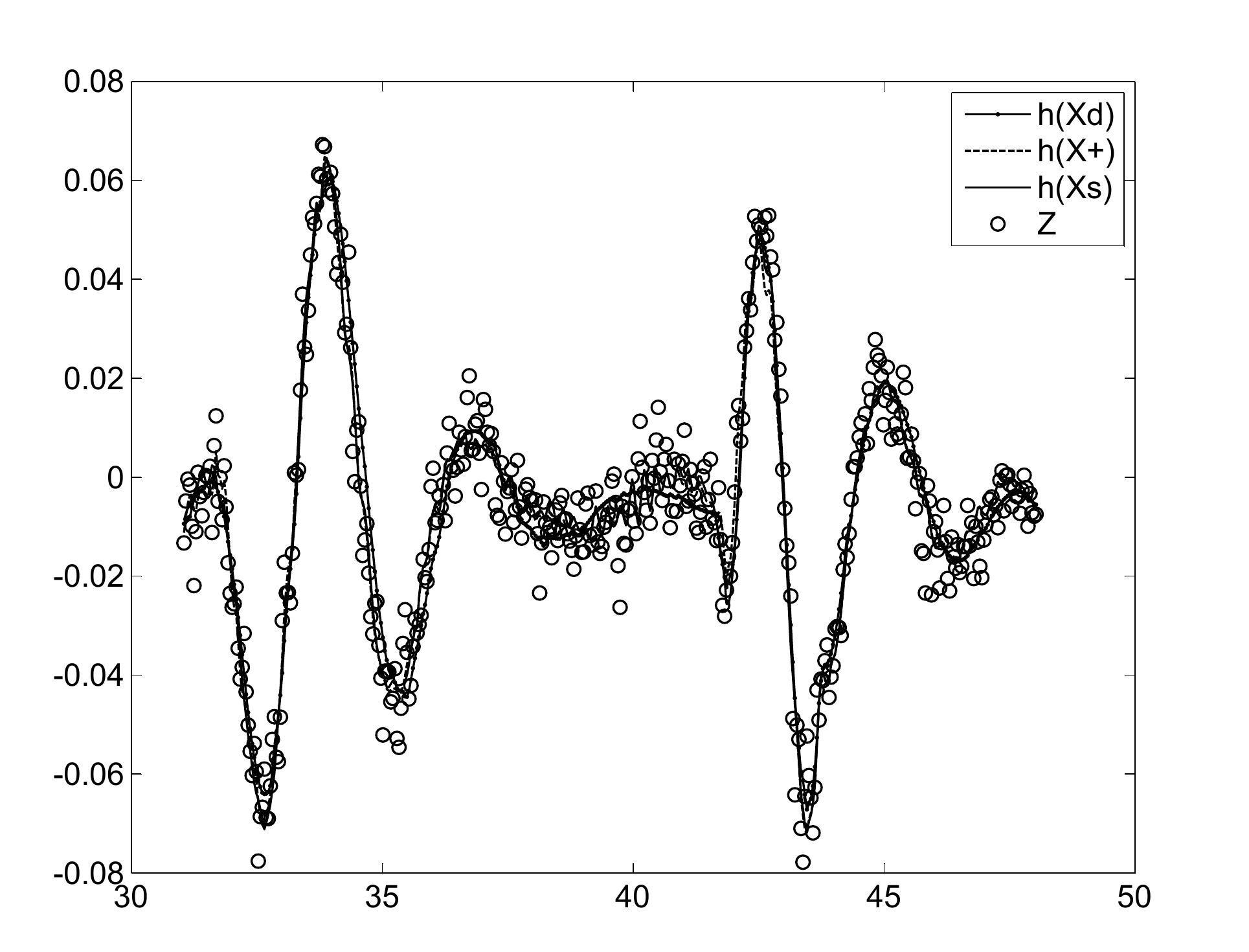}
\caption{Comparison of the predicted dynamics, posterior, smoothed}
\caption*{and the measurement in $ft/sec^2$ (lateral acceleration) vs time}
\label{realQ5_h5}
\end{figure}

 Case-3 real data is run using the reference EKF (\textbf{Q} $>$ 0) with 100 iterations. Fig.\ref{input5}-\ref{case5_alpha} are the inputs used in state equations. The Fig.\ref{realQ5_P0} shows the variation of parameter estimates and its initial covariance $\mathbf{P_0}$ with iterations and a similar Fig.\ref{realQ5_R} for \textbf{Q} and \textbf{R}. The values of \textbf{J1-J3} are close to the number of measurements ($m=5$) with \textbf{J6-J8} are close to the number of states ($n=4$) as shown in Fig.\ref{realQ5_J} and Table-\ref{tbcase5Q}. The \textbf{J5} is the negative log likelihood cost function. The later Fig.\ref{realQ5_s1}-\ref{realQ5_h5} compares (i) the state dynamics based on the estimated parameter after the filter pass through the data, (ii) the state after measurement update, (iii) the smoothed state and (iv) the measurement. Unlike in the simulated studies the estimated measurement and process noise did not have constant statistical characteristics across time. Another experiment was carried out by generating a typical data set by using the estimated theta and injecting the estimated \textbf{Q} and \textbf{R} as additive white Gaussian noise. This is to determine the effect of non white and non Gaussian noise distribution in the real data on the CRBs. After each iteration in the reference recipe the $\Theta$, \textbf{Q} and \textbf{R} were reset as from the real data. Similar experiment was also conducted by updating $\Theta$ as well. It turned out that there is not much of a difference in the final estimates and the CRBs. Two other filter runs were carried out using the MT and MS statistics for the estimation of \textbf{Q} and \textbf{R} with scaled up $\mathbf{P_0}$. The behaviour of the various cost function and in particular \textbf{J6} and \textbf{J7} in Table-\ref{tbcase5QMTMS} shows that the choice of the filter statistics for estimating \textbf{Q} and \textbf{R} in the proposed reference approach is the best possible when compared to other approaches. 

\subsection{Remarks on Case - 3}
\begin{figure}[h]
\includegraphics[width=6in,height=3in]{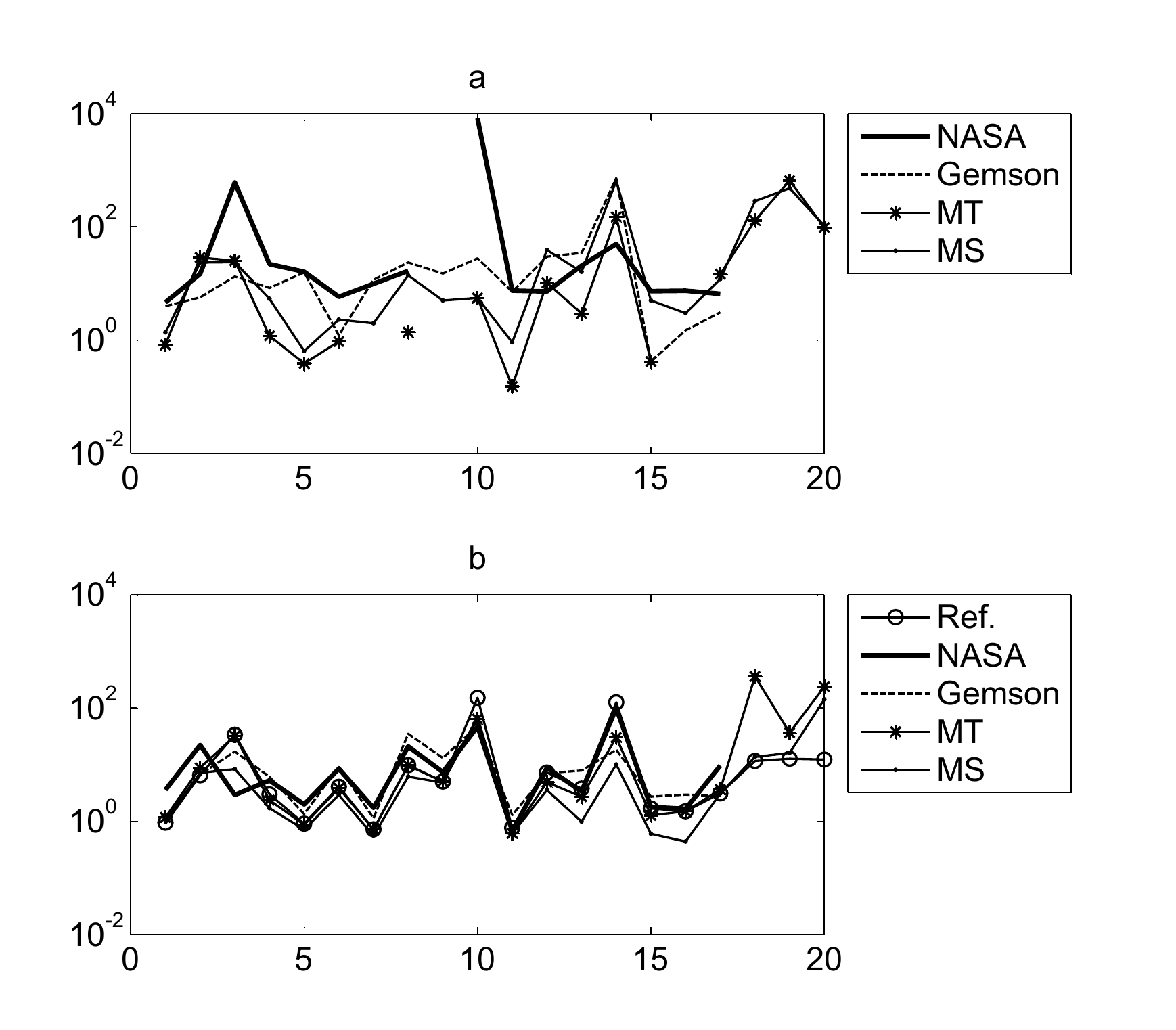}
\caption{Comparison of (a) Absolute percentage error with respect to the reference of the parameter estimates and (b) $\%$CRBs  by different methods}
\label{comp5}
\end{figure}

The NASA results have been generated assuming \textbf{Q} = 0 and are comparable with reference procedure for the parameter estimates and their CRBs. Further the MT and MS methods give quite different estimates for the \textbf{Q} and \textbf{R} values than in the reference case. We believe that the reference procedure provides the best possible parameter estimates and their uncertainties. From the plot of \% error in the parameter estimate with respect to the reference value and the \%CRB in Fig \ref{comp5}, it can be seen that the parameters $\beta_0$, $\phi_0$, $C_{N_{\delta_a}}$, $C_{Y_p}$ and $C_{Y_r}$ are relatively weak when compared to other parameters. The CRBs as estimated by different methods generally appear to vary widely. However what is interesting is that even the estimate of the strong parameter such as $C_{L_{p}}$ varies widely among the methods. Such a behaviour of the filter across the parameter estimates shows how important is the tuning of the filter statistics namely $\mathbf{P_0}$, \textbf{Q} and \textbf{R} in parameter estimation and their uncertainties. The rounded off 100$\times$C matrix of the parameter estimates for this case is

\begin{center}
\resizebox{\textwidth}{!}{\begin{tabu}{|c|c|c|c|c|c|c|c|c|c|c|c|c|c|c|c|c|c|c|c|c|c|c|}
\hline

$\Theta$ &  $\beta_0$ &  $\phi_0$ & $C_{L_{\beta}}$ & $C_{L_{p}}$ & $C_{L_r}$ & $C_{L_{\delta_a}}$ & $C_{L_{\delta_r}}$ & $C_{L_{0}}$ & $C_{N_{\beta}}$ & $C_{N_{p}}$ & $C_{N_r}$ & $C_{N_{\delta_a}}$ & $C_{N_{\delta_r}}$ & $C_{N_0}$ &  $C_{Y_{\beta}}$ & $C_{Y_p}$ &  $C_{Y_r}$ &  $C_{Y_{\delta_a}}$  & $C_{Y_{\delta_r}}$ & $C_{Y_0}$ \\ \hline

$\beta_0$ & 100	&	-1	&	0	&	0	&	0	&	0	&	1	&	0	&	0	&	0	&	2	&	0	&	4	&	 -8	&	3	&	0	&	-9	&	-1	&	-15	&	16	\\ \hline
$\phi_0$ & -1	&	100	&	0	&	0	&	0	&	0	&	0	&	0	&	0	&	0	&	0	&	0	& 0	&	0	&	0	&	0	&	0	&	0	&	0	&	0	\\ \hline \tabucline[1.5pt]{4-9}
$C_{L_{\beta}}$ &  0	&	0	&	\multicolumn{1}{!{\vrule width 2pt}c|}{100}	&	17	&	12	&	-11	&	3	&	12	&	\multicolumn{1}{!{\vrule width 2pt}c|}{-28}	&	-5	&	-3	&	3	&	-1	& -3 &		-6	&	-1	&	-1	&	1	&	0	&	-1	\\ \hline
$C_{L_{p}}$ &  0	&	0	&	\multicolumn{1}{!{\vrule width 2pt}c|}{17}	&	100	&	3	&	-84	&	8	&	1	&	\multicolumn{1}{!{\vrule width 2pt}c|}{-5}	&	-29	&	-1	&	24	&  -2	&	0	&	-1	&	-5	&	0	&	4	&	0	&	0	\\ \hline
$C_{L_r}$ & 0	&	0	&	\multicolumn{1}{!{\vrule width 2pt}c|}{12}	&	3	&	100	&	-2	&	62	&	-57	&	\multicolumn{1}{!{\vrule width 2pt}c|}{-3}	&	-1	&	-28	&	0	& -17	&	16	&	-1	&	0	&	-6	&	0	&	-4	&	3	\\ \hline
$C_{L_{\delta_a}}$ & 0	&	0	&	\multicolumn{1}{!{\vrule width 2pt}c|}{-11}	&	-84	&	-2	&	100	&	-3	&	-6	&	\multicolumn{1}{!{\vrule width 2pt}c|}{3}	&	24	&	0	&	-29	&	1	&	2	&	1	&	4	&	0	&	-5	&	0	&	0	\\ \hline
$C_{L_{\delta_r}}$ & 1	&	0	&	\multicolumn{1}{!{\vrule width 2pt}c|}{3}	&	8	&	62	&	-3	&	100	&	-96	&	\multicolumn{1}{!{\vrule width 2pt}c|}{-1}	&	-2	&	-17	&	1	&	-29	&	27	&	0	&	0	&	-4	&	0	&	-6	&	6	\\ \hline
$C_{L_{0}}$ & 0	&	0	&	\multicolumn{1}{!{\vrule width 2pt}c|}{12}	&	1	&	-57	&	-6	&	-96	&	100	&	\multicolumn{1}{!{\vrule width 2pt}c|}{-3}	&	0	&	16	&	2	&	27	&	-29	&	-1	&	0	&	3	&	0	&	6	&	-6	\\ \hline \tabucline[1.5pt]{4-15}
$C_{N_{\beta}}$ & 0	&	0	&	-28	&	-5	&	-3	&	3	&	-1	&	-3	&	\multicolumn{1}{!{\vrule width 2pt}c|}{100}	&	17	&	12	&	-11	&	3	&	12	&	\multicolumn{1}{!{\vrule width 2pt}c|}{-27}	&	-4	&	-3	&	3	&	-1	&	-3	\\ \hline
$C_{N_{p}}$ & 0	&	0	&	-5	&	-29	&	-1	&	24	&	-2	&	0	&	\multicolumn{1}{!{\vrule width 2pt}c|}{17}	&	100	&	3	&	-84	&	8	&	1	&	\multicolumn{1}{!{\vrule width 2pt}c|}{-5}	&	-22	&	-1	&	19	&	-2	&	0	\\ \hline
$C_{N_r}$ & 2	&	0	&	-3	&	-1	&	-28	&	0	&	-17	&	16	&	\multicolumn{1}{!{\vrule width 2pt}c|}{12}	&	3	&	100	&	-2	&	62	&	-57	&	\multicolumn{1}{!{\vrule width 2pt}c|}{-3}	&	-1	&	-27	&	0	&	-16	&	15	\\ \hline
$C_{N_{\delta_a}}$ & 0	&	0	&	3	&	24	&	0	&	-29	&	1	&	2	&	\multicolumn{1}{!{\vrule width 2pt}c|}{-11}	&	-84	&	-2	&	100	&	-3	&	-6	&	\multicolumn{1}{!{\vrule width 2pt}c|}{3}	&	19	&	0	&	-22	&	1	&	2	\\ \hline
$C_{N_{\delta_r}}$ & 4	&	0	&	-1	&	-2	&	-17	&	1	&	-29	&	27	&	\multicolumn{1}{!{\vrule width 2pt}c|}{3}	&	8	&	62	&	-3	&	100	&	-96	&	\multicolumn{1}{!{\vrule width 2pt}c|}{-1}	&	-2	&	-17	&	1	&	-26	&	25	\\ \hline
$C_{N_0}$ & -8	&	0	&	-3	&	0	&	16	&	2	&	27	&	-29	&	\multicolumn{1}{!{\vrule width 2pt}c|}{12}	&	1	&	-57	&	-6	&	-96	&	100	&	\multicolumn{1}{!{\vrule width 2pt}c|}{-3}	&	0	&	15	&	1	&	25	&	-26	\\ \hline \tabucline[1.5pt]{10-21}
$C_{Y_{\beta}}$ & 3	&	0	&	-6	&	-1	&	-1	&	1	&	0	&	-1	&	-27	&	-5	&	-3	&	3	&	-1	&	-3	&	\multicolumn{1}{!{\vrule width 2pt}c|}{100}	&	21	&	7	&	-16	&	-3	&	\multicolumn{1}{c!{\vrule width 2pt}}{18}	\\ \hline
$C_{Y_p}$ & 0	&	0	&	-1	&	-5	&	0	&	4	&	0	&	0	&	-4	&	-22	&	-1	&	19	&	-2	&	0	&	\multicolumn{1}{!{\vrule width 2pt}c|}{21}	&	100	&	3	&	-90	&	9	&	\multicolumn{1}{c!{\vrule width 2pt}}{3}	\\ \hline
 $C_{Y_r}$ & -9	&	0	&	-1	&	0	&	-6	&	0	&	-4	&	3	&	-3	&	-1	&	-27	&	0	&	-17	&	15	&	\multicolumn{1}{!{\vrule width 2pt}c|}{7}	&	3	&	100	&	-2	&	64	&	\multicolumn{1}{c!{\vrule width 2pt}}{-60}	\\ \hline
$C_{Y_{\delta_a}}$  & -1	&	0	&	1	&	4	&	0	&	-5	&	0	&	0	&	3	&	19	&	0	&	-22	&	1	&	1	&	\multicolumn{1}{!{\vrule width 2pt}c|}{-16}	&	-90	&	-2	&	100	&	-5	&	\multicolumn{1}{c!{\vrule width 2pt}}{-7}	\\ \hline
$C_{Y_{\delta_r}}$ & -15	&	0	&	0	&	0	&	-4	&	0	&	-6	&	6	&	-1	&	-2	&	-16	&	1	&	-26	&	25	&	\multicolumn{1}{!{\vrule width 2pt}c|}{-3}	&	9	&	64	&	-5	&	100	&	\multicolumn{1}{c!{\vrule width 2pt}}{-96}	\\ \hline
$C_{Y_0}$ & 16	&	0	&	-1	&	0	&	3	&	0	&	6	&	-6	&	-3	&	0	&	15	&	2	&	25	&	-26	&	\multicolumn{1}{!{\vrule width 2pt}c|}{18}	&	3	&	-60	&	-7	&	-96	&	\multicolumn{1}{c!{\vrule width 2pt}}{100}	\\ \hline \tabucline[1.5pt]{16-21}

\end{tabu}}
\end{center}

\vspace{1cm}
In analyzing the correlation coefficient matrix it is firstly useful to see the combination of parameters that occur in the governing equations of the flight data analysis. The sets in the present case are the trim values ($\beta_0$ and $\Phi_0$) have very little correlation with all other estimated parameters. The parameters in the three sets ($C_{L_{\beta}}$, $C_{L_{p}}$, $C_{L_r}$, $C_{L_{\delta_a}}$, $C_{L_{\delta_r}}$, $C_{L_{0}}$), ($C_{N_{\beta}}$, $C_{N_{p}}$, $C_{N_r}$, $C_{N_{\delta_a}}$, $C_{N_{\delta_r}}$, $C_{N_{0}}$), and ($C_{Y_{\beta}}$, $C_{Y_{p}}$, $C_{Y_r}$, $C_{Y_{\delta_a}}$, $C_{Y_{\delta_r}}$, $C_{Y_{0}}$), all have very similar correlations among themselves as seen in the blocks of matrices and the reason is as follows. There is coupling of the dynamical motion due to the states and the controls. If a certain state or control is excited relatively higher than others then the estimated parameter that multiplies it will have lower correlation with other estimates in the set and vice versa. Since the parameter sets are similarly excited all of them have similar correlation coefficient matrices. This feature is similar to the spring, mass, and damping system considered in Part-1 of the paper.

\clearpage

\section{Conclusions} 

The present adaptive filtering approach based on the reference recursive recipe is applied to the more involved cases of three sets of real airplane flight test data which have a larger number of state, measurements, and unknown parameters. A closer look at the correlation coefficients that is generally ignored in such studies of estimating the unknown parameters indicates the necessity to process the data by including the process noise \textbf{Q} in addition to the measurement noise \textbf{R}. Generally the parameter estimates across the various approaches are close but their CRB can vary much more among them. In particular the generalized cost functions based on balancing the state and measurement equations using the many filter outputs introduced in the present work help to show the present approach to be better than the earlier approaches and also help to identify definitive results from deceptive results.

\section*{Acknowledgments} 
Our grateful thanks are due to Profs. R. M. Vasu (Dept. of Instrumentation and Applied Physics), D. Roy (Dept. of Civil Engineering), and M. R. Muralidharan (Supercomputer Education and Research Centre) for help in a number of ways without which this work would just not have been possible at all and also for providing computational facilities at the IISc, Bangalore.

\begin{table}[h]
\caption{Real flight test data case-3 results ($\Theta,\sigma_\Theta$).}{}
\label{tbcase5Q}
\begin{center}
\begin{tabular}{|c| c| c| c| c| c|c|c|c|c|c| }
\hline
$\Theta$ & 
\makecell{Reference} &   
\makecell{NASA} &   
\makecell{Gemson} &  
MT &
MS
\\ \hline

\makecell{$C_{Y_{\beta}}$ \\ \\ $C_{Y_{\delta_r}}$ \\ \\ $\beta_0$ \\ \\ $C_{L_{\beta}}$ \\ \\ $C_{L_{p}}$ \\ \\ $C_{L_r}$ \\ \\ $C_{L_{\delta_a}}$ \\ \\ $C_{L_{\delta_r}}$ \\ \\ $C_{L_{0}}$ \\ \\ $ \phi_0$ \\ \\ $C_{N_{\beta}}$ \\ \\ $C_{N_{p}}$ \\ \\ $C_{N_r}$ \\ \\ $C_{N_{\delta_a}}$ \\ \\ $C_{N_{\delta_r}}$ \\ \\ $C_{N_0}$ \\ \\ $C_{Y_0}$ \\ \\ $C_{Y_p}$ \\ \\ $C_{Y_r}$ \\ \\ $C_{Y_{\delta_a}}$ \\ } &


\makecell{  -0.4579 \\   (0.0043) \\   0.1040 \\   (0.0067)\\   -0.0143 \\   (0.0048) \\   -0.0168 \\   (0.0005) \\  -0.3100  \\  (0.0028) \\    0.0740 \\   (0.0030)  \\  0.0557 \\   (0.0004) \\    0.0072  \\  (0.0007) \\
  -0.0020  \\ (0.0001) \\   0.0018 \\    (0.0027)  \\  0.0656 \\    (0.0005) \\  -0.0429  \\  (0.0031) \\
   -0.0880  \\  (0.0033) \\    0.0004  \\  (0.0005) \\   -0.0478  \\  (0.0008) \\    0.0067 \\   (0.0001) \\ -0.0259  \\  (0.0008) \\   -0.2828  \\  (0.0327) \\    0.2224 \\   (0.0281) \\    0.0384 \\   (0.0047) } &

\makecell{ -0.4792 \\  (0.01711)  \\ 0.0887 \\ (0.01955)  \\ -0.10116  \\ (0.00294)  \\-0.0205 \\(0.00107) \\  -0.36  \\ (0.00713) \\  0.0697  \\ (0.005884)  \\ 0.0612 \\  (0.001050) \\ 0.006 \\ (0.001252) \\ -0.002   \\ (0.0001467)  \\ 0.1506 \\ (0.07034) \\  0.0705  \\  (0.000478)  \\ -0.046 \\  (0.004006) \\ -0.1062 \\  (0.003562) \\   0.0006 \\ (0.0005924) \\ -0.0513  \\ (0.0009139)  \\ 0.0072  \\ (0.0001181) \\ -0.0242  \\ (0.002307) \\ -- \\ -- \\ -- \\--\\-- \\--} &

\makecell{ -0.4761 \\ (0.0043)  \\  0.0981 \\ (0.0065)  \\  -0.0124  \\ (0.0021)  \\ -0.0182 \\ (0.0011) \\ -0.3585  \\  (0.0048) \\ 0.0731  \\  (0.0066)  \\0.0622 \\ (0.0007) \\ 0.0089 \\  (0.0031) \\ -0.0023   \\ (0.0003)   \\0.0023 \\  (0.0011)  \\  0.0703  \\ (0.0009)  \\ -0.0557 \\ (0.0039) \\ -0.0576 \\   (0.0045) \\ 0.0033 \\ (0.0006) \\ -0.048  \\  (0.0013)  \\ 0.0068  \\ (0.0002) \\ -0.0251  \\(0.0007)  \\ -- \\ -- \\ -- \\--\\-- \\--} &

\makecell{  -0.4541 \\  (0.0053) \\  0.0741 \\  (0.0065) \\ -0.0107  \\ (0.0034)  \\ -0.0170  \\ (0.0004) \\ -0.3112 \\  (0.0027)  \\  0.0733  \\ (0.0028) \\  0.0557\\ (0.0004) \\ 0.0073 \\ (0.0007)  \\  -0.0020   \\  (0.0001)   \\ 0.0019 \\   (0.0012) \\ 0.0657  \\ (0.0004) \\ -0.0473 \\  (0.0023)  \\ -0.0854 \\  (0.0023)  \\  0.0010 \\ (0.0003) \\ -0.0476 \\ (0.0006) \\  0.0067  \\ (0.0001)  \\-0.0221 \\ (0.0008)  \\   0.0821 \\ (0.2941)  \\ -1.2336  \\   (0.4470)   \\ 0.0011 \\ (0.0026)} &

\makecell{  -0.4642 \\ (0.0049)  \\  0.0797  \\ (0.0057) \\ -0.0109 \\ (0.0009) \\   -0.0177  \\ (0.0003) \\-0.3080  \\ (0.0022) \\ 0.0757 \\ (0.0022) \\  0.0546 \\ (0.0003) \\  0.0082 \\ (0.0005) \\  -0.0021   \\(0.0001)\\ 0.0019  \\ (0.0013) \\ 0.0662 \\ (0.0004)  \\ -0.0596 \\ (0.0021)  \\ -0.1021 \\(0.0010) \\   0.0030 \\(0.0003) \\ -0.0502  \\ (0.0003) \\ 0.0069  \\ (0.00003) \\ -0.0228   \\ (0.0007)  \\ 0.5223 \\ (0.0710)  \\ -0.8452  \\  (0.1355)   \\ -0.0017 \\ (0.0024)} 
\\ \hline 

\end{tabular}
\end{center}

\vspace{1cm}

\caption{Real flight test data case-3 results* (\textbf{R,Q,J}).}{}
\label{tbcase5QMTMS}
\begin{center}
\begin{tabular}{|c| c| c|| c| c| c|| c|c|c|c|c|c| }
\hline

\makecell{\textbf{R} \\ $\times10^{-6}$\\ (Ref)}&  
\makecell{\textbf{Q} \\ $\times10^{-6}$\\ (Ref)}&  
\makecell{\textbf{J1-J8} \\(Ref) }&   

\makecell{\textbf{R} (MT)\\ $\times10^{-6}$ }&  
\makecell{\textbf{Q} (MT)\\ $\times10^{-6}$}&  
\makecell{\textbf{J1-J8} \\(MT) }&

\makecell{\textbf{R} (MS) \\ $\times10^{-6}$}&  
\makecell{\textbf{Q} (MS)\\ $\times10^{-6}$}&  
\makecell{\textbf{J1-J8} \\(MS) }
\\ \hline

\makecell{ 0.0871 \\   0.0623 \\   0.2255 \\   0.0200 \\  43.8064} &
\makecell{ 4.2163  \\  5.1340  \\  4.9426 \\   1.4324} &
\makecell{  4.7650  \\  4.8321 \\   3.5272  \\  0.0004 \\ -55.0111  \\  3.9673  \\  3.9669  \\  3.8171} &

\makecell{ 2.83  \\  18.86  \\  3.88  \\  4.09 \\   73.6} &
\makecell{2.0481  \\  3.7876  \\  1.0057  \\  0.5502} &
\makecell{ 4.3450 \\   4.3888 \\   3.1039 \\   0.0003 \\ -51.3490  \\  9.3006  \\  9.3005 \\ 3.5105} &

\makecell{  13.03  \\  88.07 \\   4.5  \\  36.36  \\  60.13} &
\makecell{  0.0005 \\   0.0007  \\  1.0975  \\  0.0016} &
\makecell{  4.8127 \\   4.8200  \\  4.5173 \\   0.0003 \\ -47.2441 \\   7.5931  \\  7.5896 \\ 3.9681}

\\ \hline

\end{tabular}
*Cost functions are not close to their expected values in MT and MS methods.
\end{center}
\end{table}

\clearpage
\begin{small}

\begin{itemize}

\item[] \textbf{{\large References}}

\item[] Anilkumar, A. K. (2000) Application of Controlled Random Search Optimisation Technique in MMLE with Process Noise. \emph{MSc Thesis, Dept. of Aerospace Engineering IISc, Bangalore}.

\item[] Bavdekar, V. A.,  Deshpande, A. P. and  Patwardhan, S. C. (2011) Identification of process and measurement noise covariance for state and parameter estimation using extended Kalman filter. \emph{Journal of Process control, 21, pp. 585-601}.

\item[] Brown, R. and  Hwang, P. (2012) Introduction to Random Signals and Applied Kalman Filtering, With MATLAB Exercises, 4$^{th}$ Edition. \emph{John Wiley and Sons, Inc.}

\item[]  Bohn, C. (2000) Recursive Parameter Estimation for Non Linear Continuous time systems through Sensitivity-Model based Adaptive Filters. \emph{PhD Thesis, Dept. of Electrical Engineering and Information Science}.

\item[]  Gemson, R. M. O. and  Ananthasayanam, M. R. (1998) Importance of Initial State Covariance Matrix for the Parameter Estimation Using Adaptive Extended Kalman Filter. \emph{AIAA-98-4153, pp. 94-104}.

\item[] Ishimoto, S. (1997) New Algorithm Of Maximum Likelihood Parameter Estimation for Flight Vehicles. \emph{AIAA -97- 3784, pp. 791-801}.

\item[] Jategaonkar, R. V. (2006) Flight Vehicle System Identification :  A Time Domain Methodology. \emph{Published by AIAA, Volume 216.}. 

\item[]  Jazwinski, A. H. (1970) Stochastic Process and Filtering Theory. \emph{Academic Press, New York}.

\item[] Kailath, T. (1970) An Innovation Approach To Detection and Estimation Theory. \emph{Proceedings of the lEEE,  Vol. 58, Issue: 5, pp. 680 -695}.

\item[] Kalman, R. E. (1960) A New Approach to linear Filtering and Prediction Problems. \emph{Transactions of the ASME-Journal of Basic Engineering, 82 (Series D) : pp. 35-45}.

\item[] Kalman, R. E., Bucy, R. S. (1961)  New Results in Linear Filtering and Prediction Theory. \emph{J. Fluids Eng. 83(1), 95-108}.

\item[] Klein, V., (1979) Identification Evaluation Methods, Parameter Identification. \emph{AGARD-LS-104 : pp. 2-21}.

\item[] Klein, V.  and  Morelli, E.A. (2006)  Aircraft System Identification. Theory and Practice. \emph{AIAA Edu. Series}.

\item[] Ljung, L. (1979) Asymptotic behaviour of the EKF as a parameter estimator for linear systems. \emph{IEEE trans. Automatic control, Vol. AC 24, pp. 36-50}.

\item[] Maine, R. E. and Iliff, K. W. (1981) Programmer's manual for MMLE3, a general Fortran program for Maximum Likelihood parameter estimation", \emph{NASA TP-1690}.

\item[] Maine, R. E. and Iliff, K. W. (1981) Formulation and Implementation of a Practical Algorithm for Parameter Estimation with Process and Measurement Noise. \emph{SIAM J. Appl. Math., vol. 41, no. 3, pp. 558-579}.

\item[] Manika, S., Bhaswati, G. and Ratna, G. (2014) Robustness and Sensitivity Metrics for Tuning the Extended Kalman Filter. \emph{IEEE Trans. on Instrumentation and Measurement, Vol. 63, No. 4, pp. 964-971}.

\item[] Maybeck, P. S. (1979) Stochastic Models, Estimation, and Control: Volume 1, " \emph{New York: Academic Press}.

\item[]  Mohamed, A. H. and Schwarz,  K. P. (1999)  Adaptive Kalman Filtering for INS/GPS. \emph{Journal of Geodesy,  Volume 73, Issue 4, pp 193-203}.

\item[] Rauch,  H. E., Tung, F. and Striebel, C. T. (1965)  Maximum Likelihood Estimates of Linear Dynamic Systems. \emph{AIAA Journal, Vol. 3, No. 8, pp. 1445-1450}.

\item[] Schultz, G . (1976) Maximum Likelihood Identification Using Kalman Filtering Least Square Estimation. \emph{NTIS : N76-30227/2, ESA TT-258}.

\item[] Shafer, M. F. (1975) Stability and Control Derivatives of the T-37B Airplane. \emph{NASA TM X-56036}.

\item[] Shumway, R. H. and Stoffer, D. S. (1982) An approach to time series smoothing and forecasting using the EM algorithm. \emph{J. Time Series Anal., 3 : pp. 253-264}.

\item[] Shumway, R. H.,  Stoffer, D. S. (2000) Time Series Analysis and its Applications. \emph{Springer, Verlag, New York}.

\item[] Shyam, M. M., Naren Naik, Gemson, R. M. O, Ananthasayanam, M. R. (2015) Introduction to the Kalman Filter and Tuning its Statistics for Near Optimal Estimates and Cramer Rao Bound. \emph{TR/EE2015/401, Dept. of Electrical Engineering, IIT Kanpur, \url{http://arxiv.org/abs/1503.04313}}.

\item[] Stepner, D. E. and Mehra, R. K. (1973) Maximum Likelihood Identification and Optimal Input Design for Identifying Aircraft Stability and Control Derivatives. \emph{NASA Contractor Report no. NASA CR-2200}.

\end{itemize}

\end{small}
\end{document}